\documentclass[12pt]{article}
\usepackage{palatino}
\usepackage{a4}
\usepackage{axodraw}
\usepackage{cite}
\usepackage{graphicx}
\usepackage{amssymb}
\usepackage{amsmath}
\usepackage{epsfig}
\usepackage[tmargin=1cm,bmargin=1cm,lmargin=2cm,rmargin=2cm]{geometry}

\newcommand{\qb}{\ensuremath{\overline q}}

\newcommand{\be}{\begin{equation}}
\newcommand{\bdm}{\begin{displaymath}}
\newcommand{\bea}{\begin{eqnarray}}
\newcommand{\beastar}{\begin{eqnarray*}}

\newcommand{\ds}{\ensuremath{\delta_s}}
\newcommand{\dc}{\ensuremath{\delta_c}}

\newcommand{\ee}{\end{equation}}

\newcommand{\e}{\ensuremath{\epsilon}}

\newcommand{\edm}{\end{displaymath}}
\newcommand{\eea}{\end{eqnarray}}
\newcommand{\eeastar}{\end{eqnarray*}}

\newcommand{\mur}{\ensuremath{\mu_{R}}}
\newcommand{\muf}{\ensuremath{\mu_{F}}}
\newcommand{\Msq}{\ensuremath{ { \overline {|M^{(0)}|^2} }}}

\newcommand{\rarrow}{\ensuremath{\rightarrow}}
\newcommand{\ReDs}{\ensuremath{{\cal R}e {\cal D}_s}}
\newcommand{\ImDs}{\ensuremath{{\cal I}m {\cal D}_s}}

\newcommand{\s}{\ensuremath{\sigma}}

\newcommand{\cd}{\ensuremath{\mathcal D}}

\newcommand{\cl}{\ensuremath{\mathcal L}}

\newcommand{\ct}{\ensuremath{\mathcal T}}

\newcommand{\m}{\ensuremath{\mu}}
\newcommand{\n}{\ensuremath{\nu}}

\nopagebreak

\begin{document}
\baselineskip 22pt

\begin{flushright}
\end{flushright}
\vskip 65pt
\begin{center}
{\large \bf
$Z$ boson pair production at the LHC to ${\cal O}(\alpha_s)$ \\in TeV scale gravity models
} 
\\
\vspace{8mm}
{\bf
Neelima Agarwal$^a$
\footnote{neel1dph@gmail.com},
}
{\bf
V. Ravindran$^b$
\footnote{ravindra@hri.res.in},
Vivek Kumar Tiwari$^a$
\footnote{vivekkrt@gmail.com},
Anurag Tripathi$^b$
\footnote{anurag@hri.res.in}
}\\
\end{center}
\vspace{10pt}
\begin{flushleft}
{\it
a)~~ Department of Physics, University of Allahabad, Allahabad 211002, India. \\
b)~~Regional Centre for Accelerator-based Particle Physics,\\
~~~~Harish-Chandra Research Institute,
 Chhatnag Road, Jhunsi, Allahabad 211019, India.\\
}
\end{flushleft}
\vspace{10pt}
\centerline{\bf Abstract}
The first results on next-to-leading order QCD corrections to production of two $Z$ bosons  
in hadronic collisions in the large extra dimension ADD model are presented.
Various kinematical distributions are obtained to order $\alpha_s$ in QCD by 
taking into account all the parton level subprocesses.  
We estimate the impact of the QCD corrections on various observables and find that they
are significant.  We also show the reduction in factorization scale uncertainty when
${\cal O}(\alpha_s)$ effects are included.

\vskip12pt
\vfill
\clearpage

\setcounter{page}{1}
\pagestyle{plain}

\section{Introduction}
\newcommand{\vn}{\ensuremath{{\vec n}}}
The Large Hadron Collider (LHC) which will operate at an enormous
center of mass energy (${\sqrt S}=14 TeV$) offers to shed light on
the mechanism of spontaneous symmetry breaking.  It also promises
to be a discovery machine and it is hoped that some signals of 
new physics beyond the standard model (SM) will be observed.
Many exciting possibilities have been envisaged; most popular ones
are supersymmetry, a symmetry which relates fermions to bosons, 
and the possibility of extra dimensions. 
In this paper we will consider
the large extra dimension model by Arkani-Hamed, Dimopoulos and Dvali
(ADD) \cite{ArkaniHamed:1998rs, Antoniadis:1998ig}.
There are many important discovery channels at the LHC
such as $\gamma \gamma, ZZ, W^+W^-$, jet production etc.  
These processes have already been studied in the context  
of anomalous triple gauge boson vertices \cite{Hagiwara:1986vm}.
In the SM the production of two $Z$ bosons is suppressed as it begins at the 
order $e^4$ in the the electromagnetic coupling and also because
of the large $ZZ$ production threshold.  
The two $Z$ bosons can couple to Kaluza Klein $(KK)$ gravitons, thus $ZZ$ pairs can be 
produced through virtual graviton exchange at the leading order. These
observations make $ZZ$ production one of the important discovery channels.
At LHC, Quantum Chromodynamics (QCD) plays an important role through
higher order radiative corrections in reducing various
theoretical uncertainties in the prediction of various important observables
to understand both SM as well as new physics scenarios. 
In this paper we will consider production of $Z$ boson pairs at the LHC 
at next-to-leading order (NLO) accuracy in the strong coupling constant.

Leading order studies for $ZZ$ production in the SM can be found
in \cite{Brown:1978mq}.  $Z$ pair with a large transverse momentum jet 
at LO was studied in \cite{Baur:1988cq}.
LO studies for $ZZ$ production in the context of extra-dimension models 
have already been carried out 
in \cite{Kober:2007bc ,Gao:2009eq}
and limits on the parameters of the model have been placed. But, as
we know a leading order (LO) computation is very sensitive to the choice
of factorization scale ($\mu_F$) which enters through the parton distribution
functions at this order. To have a precise prediction  of cross section 
and to have tighter constraints on the parameters of the model it is 
necessary to carry out a full NLO calculation in the strong coupling constant.
This reduces the sensitivity of observables to the factorization scale $\mu_F$.
Because of its importance $ZZ$ production has also been studied to NLO accuracy
in the SM \cite{Ohnemus:1990za ,Mele:1990bq}.
These results were subsequently updated in \cite{Campbell:1999ah ,Dixon:1999di}.
These studies provide the precise estimate of higher order effects through $K$ factor 
as well as the sensitivity of the predictions to factorization scale.  Importantly,
the corrections turned out to be larger than the expectations based on
soft gluon effects justifying a full-fledged NLO computation taking into
all the processes. 
The significance of NLO computations in the extra dimension models for the 
two photon production \cite{Kumar:2008pk ,Kumar:2009nn}
and Drell-Yan production \cite{Mathews:2004xp,Kumar:2006id}
has already been demonstrated.
Although NLO results are available in SM, they do not exist in literature in 
the context of ADD model for $Z$ boson pair production, which is the material of the present paper.

The results which are presented in this paper are obtained using our 
NLO Monte Carlo code (which is implemented on FORTRAN 77) that can easily
accommodate any cuts on the final state particles and obtain various 
kinematical distributions. Our code is based on the method of two cutoff
phase space slicing (for a review of the method see \cite{Harris:2001sx}) 
to deal with soft and
collinear singularities in the real emission contributions. The idea is to 
separate the phase space into soft and collinear regions using small 
dimensionless parameters $\delta_s$ and $\delta_c$ respectively. These singularities
that appear as poles in $\epsilon$ in dimensional regularization $(n=4+\epsilon)$ in 
the soft and collinear regions either cancel with virtual contributions or 
are mass factorized into parton distributions functions.   We have evaluated
all the soft and collinear sensitive pieces coming from real emission and virtual 
processes analytically.  Traces of Dirac gamma matrices and reduction
of one-loop tensor integrals that appear in the intermediate stages of the computation
are systematically performed using
the symbolic manipulation program FORM \cite{Vermaseren:2000nd}.  The $\gamma_5$ matrices
that appear in the intermediate stages of the computation require special care
as they are not defined in arbitrary dimensions.   We have used 
naive anti-commutation relations between
$\gamma_5$ and other gamma matrices in $n$ dimensions and the resulting  
traces are then computed in $n$ dimensions as they are free of $\gamma_5$.
Alternatively, one can use other method 
namely HVBM-scheme which was proposed in \cite{'tHooft:1972fi} and generalized in \cite{BM}.
In this approach, Gamma matrices and momenta in the loop and final state phase space integrals
are split into a 4 and an $n-4$ dimensional part.  The $\gamma_5$ anti-commutes in 4 dimensions
and commutes in $n-4$ dimensions with rest of the $\gamma$ matrices.
The results obtained this way coincide with those obtained using 
the naive anti-commutation relations (see \cite{Ohnemus:1990za ,Mele:1990bq}). 
The finite pieces 
are evaluated using Monte Carlo integration. We will discuss this method briefly in a 
later section which will also serve to introduce some of the required notations.

The paper is organized as follows. First we very briefly introduce the ADD model
and set up the notations which we use in this paper. Next we discuss the calculation
in the framework of slicing method and present the analytical results, and finally
we present the kinematical distributions.

\section{Next-to-leading order computation}

The fact that electroweak symmetry breaking scale of standard model (SM) cannot be
made stable against quantum corrections within the framework of SM, leads to exciting
possibility of new physics emerging at $TeV$ scales which stabilize the electroweak scale.
One of the very fascinating possibilities is the existence of ($d$) extra spatial dimensions.
ADD introduced a model \cite{ArkaniHamed:1998rs, Antoniadis:1998ig} in which the SM fields
are localized on a 3-brane and gravity propagates in all the $4+d$ dimensions as it is
the dynamics of spacetime itself. In a simple setting all the $d$ extra dimensions are
compactified over the same scale $R$. Although the SM fields are localized on a 3-brane,
they can feel the extra dimensions through their interaction with the tower of massive
Kaluza Klein ($KK$) gravitons. The interaction Lagrangian of SM fields with the $KK$
gravitons is given by 
\cite{Giudice:1998ck, Han:1998sg}
\be
\cl = - \frac{\kappa}{2} \sum_{{\vec n}=0}^{\infty}T^{\m \n}(x) h_{\m\n}^{\vec n}(x)
\ee
where $\kappa = \sqrt{16 \pi}/M_{Pl}$
and the massive KK gravitons are labeled by a $d$-dimensional vector of positive
integers, ${\vec n}=(n_1,n_2,\ldots,n_d)$. $T^{\mu \nu}$ denotes the energy momentum
tensor of the SM.
The zero mode corresponds to the usual 4-dimension massless graviton.
For a given KK level ${\vec n}$, there are, one spin-2 state, $(n-1)$ spin-1
states, and $n(n-1)/2$ spin-0 states, and they are all mass degenerate:
\be
m_{\vec n}^2 = \frac{4 \pi {\vec n}^2}{R} .
\ee
In this paper we will consider only spin-2 KK states.
Let us denote by $M_s$ the fundamental scale in $4+d$ dimensions and
following \cite{Han:1998sg} we define the relation among the gravitational
coupling, the volume of the extra dimensions, and the fundamental
scale as
\begin{equation}
\kappa^2 R^d= 8\pi (4\pi)^{d/2}\Gamma(d/2)M_s^{-(d+2)}.
\label{mass}
\end{equation}
Although the coupling $\kappa$ is $M_{Pl}$ suppressed, the fact that there are a 
large number of Kaluza Klein (KK) modes that couple to the SM fields makes the 
cumulative effect significant. Let us denote the sum of spin-2 KK graviton propagators
by $\cd_{\rm eff}$, then $\cd_{\rm eff}$ times square of the coupling can be written as
\be
\kappa^2 \cd_{\rm eff}(s) =  \frac{8 \pi}{M_s^4} \left( \frac{\sqrt s}{M_s} \right)^{d-2}
                              \left[ \pi + 2 i I(\Lambda/{\sqrt s}) \right]
\ee
The function $I(\Lambda /\sqrt s)$ depends on the ultraviolet cutoff $\Lambda$ on the
KK modes and its expression can be found in 
\cite{Han:1998sg}. We will identify $\Lambda$ with the fundamental scale $M_s$.
The vertex Feynman rules for 
coupling of SM fields to KK gravitons can be found in 
\cite{Giudice:1998ck, Han:1998sg ,Mathews:2004xp}.
With all the relevant parameters defined and propagators and vertices known
we are equipped to carry out the calculation.

As the gravitons couple to $Z$ bosons, $PP \rarrow ZZ$ can now also proceed through 
a process where  gravitons appear at the propagator level. These new channels 
makes it possible to observe deviations from SM predictions if extra dimensions
exist. In the following we will consider gravitons only at the propagator level
and investigate this process at NLO level.

$PP \rarrow ZZ$ at NLO has three pieces of computation. A LO piece which is a
$2 \rarrow 2$ parton level process; second is the $2 \rarrow 2$ order $a_s (\equiv g_s^2/16 \pi^2)$ piece which originates from loop corrections; the third and final part 
originates from real emission process where in addition to two $Z$ bosons, a parton is
also emitted in the final state. Let us take up these three pieces in turn.

\subsection{Leading order}
A leading order parton level process has the generic form
\be
a(p_1) + b(p_2) \rarrow Z(p_3) + Z(p_4).
\ee
In SM this proceeds through quark anti-quark annihilation to two
massive $Z$ bosons as shown in Fig.~9. The coupling of fermions to $Z$ bosons is
\be
-i \frac{e \ct_W}{2} \gamma^{\mu} \left(C_v -C_a \gamma^5 \right) ,
\ee
where the coefficients $C_v$ and $C_a$ are defined as
\bea
 C_v = T_3^f - 2~ Q_f \sin^2\theta_W,  \quad \quad C_a = T_3^f,  \quad \quad
{\cal T}_W = \frac{1}{\sin\theta_W \cos \theta_W}.
\eea
The $Q_f$ and $T_3^f$ denote the electric charge and the third component of 
the weak isospin of the fermion $f$, and $\theta_W$ is the weak mixing angle.
We give below the matrix element squares summed (averaged)
over the final (initial) sate spins, colors and polarizations.
The SM at  LO gives order $e^4$ contribution to the cross sections as given 
in eq.~(\ref{Msmlo}). In addition,
two more processes are allowed as the KK gravitons can appear at the 
propagator level, $q\qb \rarrow G^* \rarrow ZZ$ and $gg \rarrow G^* \rarrow ZZ$, as 
shown in Fig.~\ref{bsmborn}. These $q \qb$ and $gg$ initiated contributions which are of 
order $\kappa^4$ are given in eqs.~(\ref{Mbsmqlo},\ref{Mbsmglo}). 

\bea
\label{Msmlo}
\Msq_{q \qb, sm} &=&  \frac{1}{64 N t^2 u^2}
                      \left( C_v^4+6~C_v^2C_a^2+C_a^4 \right)
                      e^4 {\cal T}_W^4
\\ \nonumber
 &&
\times  (n-2) \Big[ -8 m^2 t u (t + u)
        + (tu - m^4) (n-2) (t^2+u^2)
\\ \nonumber
&&   -2 m^4 (n-12) t u +  2 (n-4) t^2 u^2
    \Big]
\eea
\bea
\label{Mbsmqlo}
\Msq_{q\qb, gr}  &=&
 \frac{1}{64 N} |{\cal D}_s|^2 \kappa^4 \Bigg[
         n \Big\{8 m^8 - 16 m^6 (t + u) + t u (3 t^2 + 2 t u + 3 u^2)
\\ \nonumber
&&            + m^4 (9 t^2 + 30 t u + 9 u^2)
            - 2 m^2 (t^3 + 7 t^2 u + 7 t u^2 + u^3)
           \Big\}
\\[2ex] \nonumber
&&
 - \Big\{8 m^8 - 24 m^6 (t + u) + t u (7 t^2 + 10 t u + 7 u^2)
\\ \nonumber
&&           + m^4 (17 t^2 + 62 t u + 17 u^2)
           - 4 m^2 (t^3 + 9 t^2 u + 9 t u^2 + u^3) \Big\}
        \Bigg]
\eea
\bea
\label{Mbsmglo}
\Msq_{gg,gr} &=&
\frac{|{\cal D}_s|^2 \kappa^4}{(N^2-1)}
  \frac{1}{128} \times
 \Bigg[ 128 m^8 +9t^4 + 28 t^3 u + 54t^2 u^2 +28 tu^3 + 9u^4
\\[2ex] \nonumber
&&
       -256m^6(t+u)
       +192m^4 (t+u)^2 -64 m^2 (t+u)^3
     - \frac{72}{\left( n-1\right)^2} s^3 (4m^2-s)
\\[2ex] \nonumber
&&
     - \frac{3}{ n-1} s^2 \Big\{ 188m^4 -17t^2 -226 tu -17u^2 +60m^2(t+u) \Big\}
\\[2ex] \nonumber
&&
     + \frac{32}{(n-2)^2} \Big\{ -44m^8 +40m^6 (t+u) -40 m^2 tu(t+u) +9tu(t+u)^2
\\[2ex] \nonumber
&&
                                 + m^4(-9t^2+26tu -9u^2)
                           \Big\}
     + \frac{4}{n-2}
       \Big\{ 692 m^8 -13t^4 -196t^3u -362t^2u^2
\\[2ex] \nonumber
&&
              -196tu^3 -13u^4
              -544 m^6(t+u) -8m^4 (t^2+83tu+u^2)
\\[2ex] \nonumber
&&
            +16m^2(5t^3 +53t^2u +53tu^2 +5u^3)
       \Big\}
 \Bigg]
\eea
Next we give the interference of SM $q\qb$ process with the gravity
mediated $q\qb$ subprocess. For convenience we will denote 
$M^{(0)}_{q\qb,sm} {M^{(0) *}_{q\qb,gr}} +c.~c.$ by  
$\Msq_{q \qb, int}$. 
\bea
\label{Mintlo}
\Msq_{q \qb, int} &=&
\frac{1} { 64 N t u }
    \left (C_v^2 + C_a^2 \right ) ( 2 {\cal R}e {\cal D}_s) e^2 \kappa^2 {\cal T}_W^2
\\ \nonumber
&&  \times
              \Bigg[   4 m^6 (n-1) (t + u)
              + 4 m^2 (3 n - 7) t u (t + u)
\\ \nonumber
&&              + t u \Big\{ (8 - 3 n) t^2 - 2 (n-4) t u + (8 - 3 n) u^2 \Big\}
\\ \nonumber
&&              - m^4 \Big\{-40 t u + n (t^2 + 22 t u + u^2) \Big\}
           \Bigg]
\eea
here sm and gr repesent contributions from standard model,
gravity and interference of SM with gravity induced processes
respectively. s, t and u  are the usual Mandelstam invariants, 
$\cd(s) =\cd_{\rm eff}/i$ and $m$ denotes the mass of $Z$ boson. $N$ denotes the number of color and these
are exact expressions in $n$ dimensions. A factor of 1/2
has been included for identical  final state Z bosons.

The parton level cross sections obtained from the leading order matrix
elements are independent of factorization scale $\mu_F$. A LO hadronic
cross section obtains its dependence on $\mu_F$ solely from the parton 
distribution functions. Due to this sensitivity to $\mu_F$ the LO
predictions are generally regarded as first approximation. To have  a
result which is less sensitive to $\mu_F$ and which also includes 
missing higher order pieces we need to go beyond the leading order.

\subsection{Next-to-leading order}
In Fig.~\ref{smvrt}, the order $a_s$ loop diagrams that appear in
SM and in Fig.~\ref{bsmvrt} the diagrams that have a graviton propagator are presented.
Here we consider only 5 flavors of quarks and treat them as massless.
These diagrams contribute through their interference with the leading order
diagrams. In general loop diagrams give ultraviolet divergences and infrared
divergences when the integration over loop momenta is carried out. We
use dimensional regularization $(n=4+\epsilon)$ to regulate these divergences;
these divergences then appear as poles in $\epsilon$. Note however that
owing to the gauge invariance and the fact that the KK gravitons couple to 
SM energy momentum tensor, a conserved quantity, this process is UV finite.
Various tensor integrals were reduced to scalar integrals following the procedure
of Passarino-Veltman \cite{Passarino:1978jh}. The 4-point scalar integrals that appear in the 
$gg$ initiated {\it box} diagrams were taken from 
\cite{Duplancic:2000sk ,Bern:1993kr}.
The one loop matrix elements are recorded below. The finite pieces of matrix element
squares denoted by a superscript {\it fin} are given in the appendix.

The SM contribution is found to be
\bea
\overline{|M^{V}|^2}_{q \qb,sm}
  &=&
       a_s(\mu_R^2) f(\e,\mu_R^2,s) C_F ~\Bigg[~
       \Upsilon \left(\epsilon \right)
      ~
            {\Msq}_{q \qb, sm}
          + \overline{|M^{V}|^2}^{fin}_{q \qb,sm}
                    \Bigg],
\eea
the interference contributions of SM with the gravity mediated processes are
\bea
\overline{|M^{V}|^2}_{q \qb,int}
  &=&
       a_s(\mu_R^2) f(\e,\mu_R^2,s) C_F ~\Bigg[~
       \Upsilon \left({\e} \right)
           {\Msq}_{q \qb, int} + \overline{|M^{V}|^2}^{fin}_{q \qb,int}
 \Bigg]
\\[2ex]
\overline{|M^{V}|^2}_{gg,int}
  &=&
       a_s(\mu_R^2){C_A }  ~\Bigg[~
            \overline{|M^{V}|^2}^{fin}_{gg,int}   \Bigg]\,,
\eea
and the pure gravity contributions are
\bea
\overline{|M^{V}|^2}_{q \qb,gr}
  &=&
       a_s(\mu_R^2) f(\e,\mu_R^2,s) C_F ~\Bigg[~
       \Upsilon \left({\e} \right)\Msq_{q\qb, gr}
      +  4(2\zeta(2) -5)  \Msq_{q\qb, gr}
       ~ \Bigg]
\\ [2ex]
\overline{|M^{V}|^2}_{gg,gr}
  &=&
       a_s(\mu_R^2) f(\e,\mu_R^2,s) C_A ~\Bigg[~
 \left\{-\frac{16}{\e^2} + \frac{4}{C_A \e}  \left( {11 \over 3 } C_A -\frac{4}{3}n_f T_f\right)\right\}
{\Msq}_{gg, gr}
\nonumber \\[1ex]
              &+& \frac{1}{9} \left( 72 \zeta(2) + 70 \frac{n_f T_f}{C_A} -203  \right)
{\Msq}_{gg, gr}
\Bigg]
\eea
where
\bea
\Upsilon \left({\e} \right)
         &=& -~\frac{16}{\displaystyle{\e^2}} + \frac{12}{\displaystyle{\e}}  ,\quad \quad \quad
f(\e,\mu_R^2,s)  = \frac{ \Gamma\left( 1+{\displaystyle {\e\over 2}}\right)} {\Gamma(1+\e)}
             \left( \frac{s}{4 \pi \mu_R^2} \right)^{\frac{\displaystyle{\e}}{2}}
\eea
The theory is renormalized at scale $\mu_R$. $C_F$ is the Casimir of the fundamental representation
while $C_A$ is the Casimir of adjoint representation in the color group.
\bea
\label{casimir}
C_F &=& \frac{N^2-1}{2N} , \quad \quad C_A = N ,\quad \quad T_f = \frac{1}{2} 
\eea
$N$ is the color degree of freedom for quarks and $N^2-1$ for gluons. We can now write the 
order $a_s(\mu_R^2)$ contributions coming from virtual diagrams as,
%
\bea
\label{xvirt}
d \s^{virt} &=& a_s(\mur^2) dx_1dx_2 f(\e, \mur^2,s)
\nonumber \\[2ex]
           && \times
              \Bigg[ C_F \left(-\frac{16}{\e^2} + \frac{12}{\e} \right)
              \sum_i  d\s^{(0)}_{q_i\qb_i}(x_1,x_2,\e)  \left( f_{q_i}(x_1)f_{\qb_i}(x_2) + x_1 \leftrightarrow x_2 \right)
\nonumber \\[2ex]
           &&    + C_A \left\{-\frac{16}{\e^2} + \frac{4}{C_A\e} \left( \frac{11}{3}C_A -\frac{4}{3}n_fT_F \right) \right
\}d \s^{(0)}_{gg}(x_1,x_2,\e) \Big(f_g(x_1)f_g(x_2) \Big)
\nonumber \\[2ex]
&& + C_F \sum_i  d\s^{V,fin}_{q_i\qb_i}(x_1,x_2,\e)  \left( f_{q_i}(x_1)f_{\qb_i}(x_2) + x_1 \leftrightarrow x_2 \right)
\nonumber \\[2ex]
&& + C_A ~d\s^{V,fin}_{gg}(x_1,x_2,\e) (f_g(x_1)f_g(x_2))
\Bigg]
\eea
Note the appearance
of poles of order 2 in $\epsilon$ in the one loop matrix elements. These 
correspond to the configurations which are both soft and collinear simultaneously.
These double poles cancel when real emission contributions are included, the
remaining simple poles do not cancel completely and are factorized into the 
bare parton distribution functions at the scale $\mu_F$.
Several checks ensure the correctness of the matrix elements. The $Z$ boson
polarization sum $-g_{\mu \nu} + k_\mu k_\nu / m^2$ does not give rise to 
negative powers of $m$. 
Further, for gluon initiated process the gluon polarization sum is 
$-g_{\mu \nu} + (k_\mu n_\nu + k_\nu n_\mu)/k.n$ where $n$ is an
arbitrary light like vector and the results are independent of the vector
$n$.  The gauge parameter present in the graviton-gluon-gluon vertex \cite{Han:1998sg} 
does not appear in the matrix element square; this serves as yet another check.
Furthermore the SM matrix elements are in agreement with the literature
\cite{Ohnemus:1990za}.

At NLO we also have to include $2 \rarrow 3$ real emission processes.
A generic process is of the form
\be
a(p_1) + b(p_2) \rarrow Z(p_3) + Z(p_4) + c(p_5).
\ee
In Fig.~\ref{smreal} we show the $q \qb$ and $qg$ initiated real emission Feynman diagrams
which appear in SM. In addition, in the ADD model the $2 \rarrow 3$ diagrams
with graviton propagator are shown in Fig.~\ref{bsmreal}. Here all the three kinds,
$q\qb, qg, gg$ initiated subprocesses occur. The $2 \rarrow 3$ contributions to 
cross-section reveal the infrared divergences when the integral over the 
final state particles is carried out. As mentioned above, the sum of virtual and
real emission cross section is finite after mass factorization is carried out, we
present very briefly below  this in the framework of phase space slicing method. 
For more details we refer to  the review 
\cite{Harris:2001sx} and our earlier work \cite{Kumar:2009nn}.

Using two small dimensionless slicing parameters $\delta_s$ and $\delta_c$ the 
$2 \rarrow 3$ phase space is divided into soft and collinear regions. The soft
is defined as the part of phase space where the final state gluon is soft and 
has an energy less than $\delta_s {\sqrt s_{12}}/2$ in the center of mass frame 
of incoming partons.
In this region the cross section simplifies and we have
\bea
\label{xsoft}
d \s ^{soft} &\simeq&  a_s dx_1 dx_2 f(\e ,\mur^2,s)
\left(\frac{16}{\e^2} + \frac{16 \ln \ds}{\e} + 8 \ln^2 \ds \right)
\nonumber \\ [2ex]
&& \times \Bigg[ \Bigg( C_F~\sum_i d\s^{(0)}_{q_i\qb_i}(x_1,x_2,\e) f_{q_i}(x_1)f_{\qb_i}(x_2
)
+ x_1 \leftrightarrow x_2 \Bigg)
\nonumber \\ [2ex]
&& \quad      + C_A~ d\s^{(0)}_{gg}(x_1,x_2,\e) f_{g}(x_1)f_{g}(x_2) \Bigg] .
\eea
The region complementary to the soft region is hard region and contains 
collinear singularities. This region is thus further divided into hard 
collinear region (the region of phase space where the final state parton
is collinear to one of the initial state parton) which contains collinear
singularities and hard non-collinear region which is free of any singularities.
After mass factorization in ${ \overline{MS}}$ scheme, hard collinear region 
gives the following contribution to the cross section. 
\bea
\label{xcoll}
\!\!\!\!\!\!  d \sigma^{HC + CT}\!\!\!  &=& \!\!\!  a_s(\mu_R^2) dx_1 dx_2 f(\e, \mu_R^2, s)
\nonumber\\[2ex]
&&  \times \Bigg[\sum_i d{\hat \sigma}^{(0)}_{q_i\qb_i}(x_1,x_2,\e) \Bigg\{ \frac{1}{2} f_{\qb_i}(x_1,\mu_F) {\tilde f}_{q_i}(x_2,\mu_F) +
               \frac{1}{2} {\tilde f}_{\qb_i}(x_1,\muf)f_{q_i}(x_2,\mu_F)
\nonumber\\[2ex]
&&           + 2 \left( -\frac{1}{\e} +\frac{1}{2} \ln \frac{p_{12}}{\muf^2} \right) A_{q \rarrow q + g}
              ~  f_{\qb_i}(x_1,\muf) f_{q_i}(x_2,\muf) + x_1 \leftrightarrow x_2  \Bigg\}
\nonumber\\[2ex]
&&+ d{\hat \sigma}^{(0)}_{gg}(x_1,x_2,\e) \Bigg\{ 2~\cdot \frac{1}{2} {\tilde f}_g(x_1,\muf) f_g(x_2,\muf)
\nonumber\\[2ex]
&& + 2 \left( -\frac{1}{\e} +\frac{1}{2} \ln \frac{p_{12}}{\muf^2} \right) A_{g \rarrow g+g}
~f_g(x_1,\muf) f_g(x_2,\muf)
\Bigg\} \Bigg].
\eea
Here terms of order $\ds$ have been dropped and 
factors of 2 appear because both the incoming partons can emit gluons.
$p_{12} = (p_1+p_2)^2$ and the other definitions used in the above equation are as follows.
\bea
A _{q \rarrow q+g} \equiv \int_{1- \delta_s}^1 \frac{dz}{z} P_{qq}(z) &=& 4C_F \left(2 \ln \ds + \frac{3}{2} \right) ,
\nonumber\\[2ex]
A_{g \rarrow g +g} \equiv  \int_{1- \delta_s}^1 \frac{dz}{z} P_{gg}(z) &=&
 \left( \frac{22}{3}C_A -\frac{8}{3}n_f T_F + 8 C_A \ln \ds \right) ,
\eea
and the function ${\tilde f_{q,g}}$ are defined by
\bea
{\tilde f}_{q}(x, \mu_F)
    &=&  \int_x^{1 -\delta_s} \frac{dz}{z} f_q \left( \frac{x}{z}, \mu_F \right) {\tilde P}_{qq}(z)
      + \int_x^{1}            \frac{dz}{z} f_g \left( \frac{x}{z}, \mu_F \right) {\tilde P}_{qg}(z) ,
\nonumber \\[2ex]
{\tilde f}_{g}(x, \mu_F)
    &=&  \int_x^{1 -\delta_s} \frac{dz}{z} f_q \left( \frac{x}{z}, \mu_F \right) {\tilde P}_{gq}(z)
      + \int_x^{1}            \frac{dz}{z} f_g \left( \frac{x}{z}, \mu_F \right) {\tilde P}_{gg}(z) ,
\eea
with
\be
{\tilde P}_{ij}(z) = P_{ij}(z) \ln \left( \delta_c \frac{1-z}{z} \frac{p_{12}}{\mu_F^2} \right) +2 P^{\prime}_{ij}(z) .
\ee
and 
\be
P_{ij}(z,\e) = P_{ij}(z)+\e P_{ij}^\prime (z)
\ee
Let us now add all the order $a_s$ pieces together; the virtual cross-section $d\sigma^{virt}$
in Eq.(~\ref{xvirt}), the
soft piece $d\sigma^{soft}$ in Eq.~(\ref{xsoft}) and the mass factorized hard collinear contribution 
$d\sigma^{HC+CT}$
as given in Eq.~(\ref{xcoll}).
We see that all the poles in $\e$ cancel in the sum
\be
d\sigma^{2-body}(\ds,\dc,\mu_F) = d\sigma^{virt}+d\sigma^{soft}(\ds,\dc)+d\sigma^{HC+CT}(\ds,\dc,\mu_F).
\ee
We have made explicit the dependence on the slicing parameters and the factorization scale
and suppressed other variables. The only order $a_s$ piece, $d\sigma^{3-body}(\ds,\dc)$, 
which remains to be included is hard non collinear; it is finite as the integration over 3-body phase space
here does not include soft and collinear regions. 
%
Thus, we need to know the phase space only in $n=4$ dimensions.
It is easy to parameterize the momenta of particles in the rest frame
of the two final state $Z$ bosons and later lorentz transform to the 
laboratory frame. We can take 
$p_1$ to define the $z$ axis, and using $p_1$ and $p_2$ 
define $y-z$ plane. 
\bea 
p_1 &=& E_1 \Big(1 ~,0, ~0, 1 \Big),
\nonumber \\[2ex]
p_2 &=& E_2 \Big(1, ~0, ~\sin \psi, ~ \cos \psi \Big),
\nonumber \\[2ex]
p_3 &=& \frac{\sqrt{p_{34}}}{2}(1,\beta_x \sin \theta_2 \sin \theta_1, \beta_x \cos \theta_2 \sin \theta_1, \beta_x \cos \theta_1 ),
\nonumber \\[2ex]
p_4 &=& \frac{\sqrt{p_{34}}}{2}(1,-\beta_x \sin \theta_2 \sin \theta_1, -\beta_x \cos \theta_2 \sin \theta_1, -\beta_x \cos \theta_1 ),
\eea
where 
\be
\beta_x = \sqrt{1-\frac{4m^2}{p_{34}}}, \quad \quad p_{34}=(p_3+p_4)^2, 
\ee
and $ \sin \psi > 0$.  
The momentum $p_5$ is determined by momentum conservation.
The three body processes are characterized by five independent
scalar quantities:
\bea
p_{12} = (p_1 +p_2)^2, \quad
p_{15} = (p_1 -p_5)^2, \quad
p_{25} = (p_2 -p_5)^2,
\nonumber \\[2ex]
p_{13} = (p_1 -p_3)^2, \quad
p_{24} = (p_2 -p_4)^2.
\eea
Of course any other set of five independent scalars
can be chosen. For convenience let us introduce the 
variables $x$ and $y$, where $x =p_{34}/p_{12}$
and $y$ is the cosine of the angle between $p_1$ and $p_5$.
We have 
\be
\frac{4 m^2}{p_{12}} \leq x \leq 1, \quad -1 \leq y \leq 1,
\ee
and
\be
p_{15} = -\frac{p_{12}}{2} (1-x)(1-y), \quad p_{25}= - \frac{p_{12}}{2}(1-x)(1+y).
\ee
The phase space is given in 4-dimensions by
\be
d \Gamma_3 = \frac{1}{(4 \pi)^2}~ \frac{\beta_x}{16 \pi} d \cos{\theta_1} dx
             \frac{p_{12}}{2 \pi} (1-x) dy d\theta_2 .
\ee
The integration over the 3-body phase space is carried out using Monte Carlo,
and it is constrained to avoid collinear and  soft regions. The $q\qb$ and $gg$
initiated processes contain both kinds of divergences so the integral
is constrained using $\ds$ and $\dc$ to avoid these regions. The $qg$ initiated 
process, however, contain only collinear singularities (as soft fermions do not
give singularities) and the 3-body integration is constrained using only $\dc$.

We want to express the momenta in the laboratory frame so as to facilitate
implementation of any cuts, such as rapidity cut, on the final state particles. 
The transformation matrix can be obtained by first 
boosting to the $p_1+p_2$ rest frame and rotating 
to align $p_1$ and $p_2$ parallel to the $z-$axis.
Finally we will boost to the laboratory frame.
Successively carrying out these transformations we obtain the 
final transformation matrix $M_{ij}$ whose components are as given 
below:
\bea
M_{00} = \frac{x_1 E_2 + x_2 E_1}{x_1 x_2 \sqrt{S}}, 
\hspace {2.7cm} \quad M_{33} = \frac{ -2E_1 (x_1 E_2 - x_2 E_1) + x_2 x_1^2 S }{2E_1 x_1 x_2 \sqrt{S}},
\eea
\bea 
M_{02} = - \frac{x_1 \sqrt{x_1 x_2 S (4E_1E_2 -x_1 x_2 S)}} {2E_1 x_1 x_2 \sqrt{S}},
\quad M_{20} = - \frac{\sqrt{x_1 x_2 S (4E_1E_2 -x_1 x_2 S)}} {x_1 x_2 S}, 
\eea
\be
M_{03} = \frac{ -2E_1 (x_1 E_2 + x_2 E_1) + x_2 x_1^2 S }{2E_1 x_1 x_2 \sqrt{S}}, 
\hspace {2.7cm}  \quad M_{30} = \frac{x_1 E_2 - x_2 E_1}{x_1 x_2 \sqrt{S}},
\ee
\bea
M_{23}=- M_{20},  \quad M_{32}=M_{02}.
\eea
The remaining  matrix elements are zero. 
$S$ denotes the center of mass energy of the colliding hadrons.
The energies $E_1$ and $E_2$ and $\cos \psi$ 
can be expressed as 
\be
\cos \psi = 1 -\frac{x_1 x_2 S}{2E_1 E_2},
\ee
\bea
E_1 = \frac{1}{4} \sqrt{ \frac {x_1 x_2 S}{x} } \Big[ 2 -(1-x)(1-y) \Big], 
\quad E_2 = \frac{1}{4}  \sqrt{ \frac {x_1 x_2 S}{x} }  \Big[ 2 -(1-x)(1+y) \Big] .
\eea
Applying the above transformation we can obtain all the momenta in the Laboratory frame
and can impose any restrictions at the Monte Carlo level. We do not give the matrix elements
for $2 \rarrow 3$ processes as the expressions are large and can be obtained on request.

The NLO result is sum $d\sigma^{LO} + d\sigma^{2-body}(\ds,\dc,\mu_F) + d\sigma^{3-body}(\ds,\dc)$.
The sum $d\sigma^{2-body}$ $(\ds,\dc,\mu_F)$ $+ d\sigma^{3-body}(\ds,\dc)$ constitutes QCD correction,
but  $d\sigma^{2-body}(\ds,\dc,\mu_F) $ and \\
$ d\sigma^{3-body}(\ds,\dc)$ independently are not
physical quantities as these depend on the (arbitrary) slicing parameters. The sum of these two 
pieces should be independent of the slicing parameters as these were introduced at the intermediate
stages of calculation. A verification of this, in the next section, will serve as a test on the code
as to the correct implementation of the phase space slicing method.  

\section{Results}
\begin{figure}[ht]
\centerline{\epsfig{file=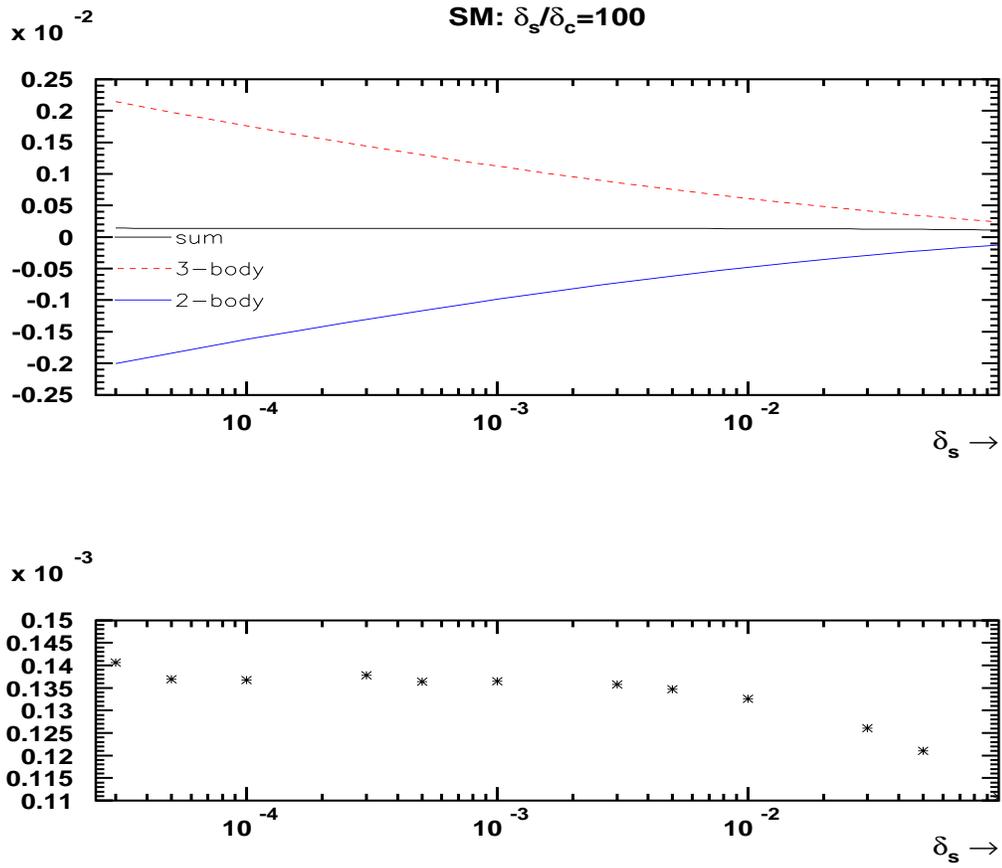,width=15cm,height=12cm,angle=0} }
\caption{Variation of 2-body and 3-body contributions (of $d\sigma /dQ$ at $Q=800~GeV$ in SM) and their sum
with $\delta_s$.
Here $\delta_s/\delta_c =100$ has been used.  }
\label{dltsm}
\end{figure}

\begin{figure}[ht]
\centerline{
\epsfig{file=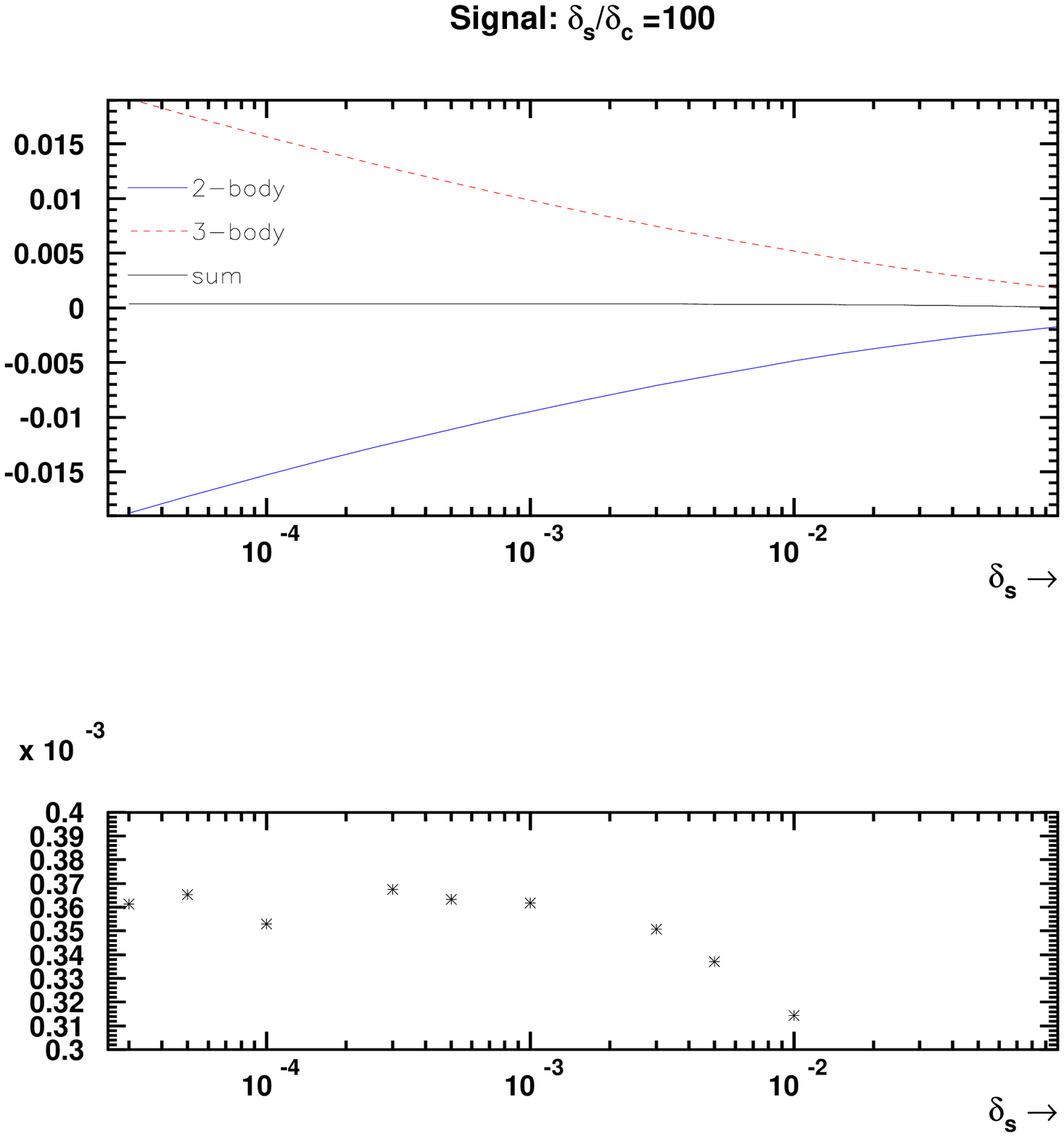,width=15cm,height=12cm,angle=0}}
\caption{Variation of 2-body and 3-body contributions (of $d\sigma /dQ$ at $Q=800 GeV$ in signal) 
and their sum with $\delta_s$.
Here $\delta_s/\delta_c =100$ has been used.  }
\label{dltsig}
\end{figure}

In the previous section we have given all the relevant analytical results, now we proceed 
to determine some kinematical distributions. First we demonstrate that the sum of 2-body and
3-body contributions is fairly independent of the slicing parameters. In Fig.~\ref{dltsm} (for SM) and 
Fig.~\ref{dltsig} (for signal) we show the variations of these two pieces with the slicing parameters in 
invariant mass, $Q=\sqrt{(p_3 + p_4)^2}$, distribution at a value of invariant mass 
equal to $800 GeV$. Here both $\ds$ and $\dc$
are varied together with the ratio $\ds/\dc$ fixed at a value of 100 
\cite{Harris:2001sx}. We note that the sum of 
2-body and 3-body contributions is fairly stable against variations in these parameters and this gives us confidence in our code.
In what follows we will use $\delta_s =10^{-3}$ and $\delta_c=10^{-5}$.

Below we present various distributions for the LHC with a center of mass energy of $14~TeV$ as a 
default choice. However we will also present some results for a center of mass energy of $10~TeV$ for the LHC.
For numerical evaluation,
the following SM parameters 
\cite{Amsler:2008zzb} are used
\be 
m = 91.1876~ GeV,  \quad \sin^2 \theta_W = 0.231
\ee
where $\theta_W$ is the weak mixing angle.
For the electromagnetic coupling constant $\alpha$ we use $ \alpha^{-1} = 128.89$. 
CTEQ6 \cite{Pumplin:2002vw, Stump:2003yu} density sets are used for parton distribution 
functions. 2-loop running for the strong coupling constant is used which is given by the 
expression,
\be
\label{asNLO}
a_s(\mu_R^2) = \frac{1}{\beta_0 \ln {\displaystyle  \frac{\mu_R^2}{\Lambda_{QCD}^2} } }
            \left[ 1 - \frac{(\beta_1/ \beta_0) ~ \ln \ln {\displaystyle  \frac{\mu_R^2}{\Lambda_{QCD}^2} }  }
                            {\beta_0 \ln {\displaystyle  \frac{\mu_R^2}{\Lambda_{QCD}^2} } } \right]  .
\ee
with
\bea 
\beta_0 &=& \frac{11}{3} C_A -\frac{4}{3}n_f T_f  ,
\nonumber \\
\beta_1 &=& \frac{34}{3}C_A^2 -\frac{4}{3} n_f T_f(3C_F +5C_A)  , 
\eea
where the symbols are as given in Eq.~\ref{casimir} with number of colors equal to three.
The number of active light-quark flavors is denoted by $n_f~(=5)$ and the value of $\Lambda_{QCD}$ is 
chosen as prescribed by the CTEQ6 density sets. At leading order, that is at order $a_s^0$,  we
use CTEQ6L1 density set ( which uses the LO running $a_s$ ) with the corresponding 
$\Lambda_{QCD}=165~MeV $. At NLO we use CTEQ6M density set ( which uses 2-loop running $a_s$ )
with the $\Lambda_{QCD}=226~MeV $; this value of $\Lambda_{QCD}$ enters into the evaluation of the 
2-loop strong coupling.
The  default choice for the renormalization and factorization scale is the identification
to the invariant mass of the $Z$ boson pair ie., $\mu_F =\mu_R =Q$. Furthermore the
$Z$ bosons will be constrained to satisfy $|y_Z| < 2.5$, where $y_Z$ is 
the rapidity of a final state $Z$ boson .

\begin{figure}[ht]
\centerline{\epsfig{file=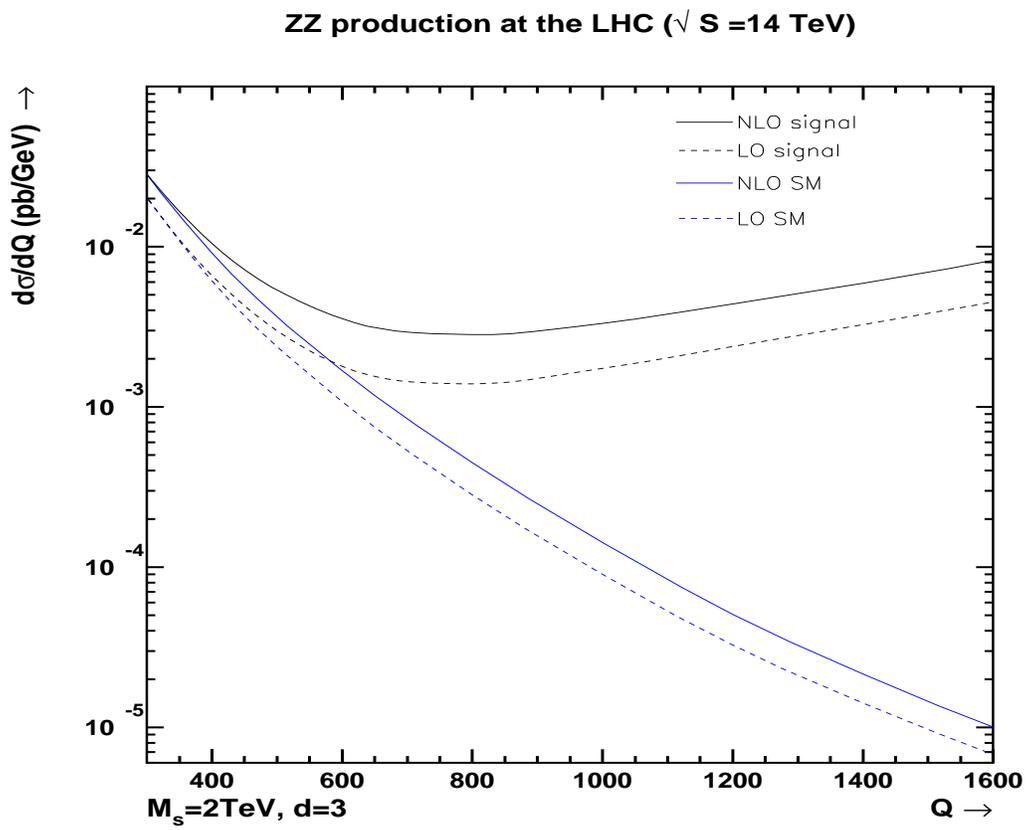,width=15cm,height=12cm,angle=0} }
\caption{Invariant mass distribution at LO and NLO in SM and for the signal at $M_s=2TeV$ and 
3 extra dimensions. }
\label{inv}
\end{figure}

In Fig.~\ref{inv} we have plotted the invariant mass distribution both for the SM and the 
signal, in the range 300 $GeV$ to 1600 $GeV$. 
The distribution is presented at the higher values of $Q$ as it is in this
region the deviations from SM are more pronounced. In this plot we display
for three extra dimensions ie., $d=3$, for fundamental scale equal to $2~TeV$. 
We see that for this choice of parameters the signal starts deviating significantly from the 
SM predictions around 400 $GeV$.
To highlight the importance of
QCD corrections we have also displayed the LO results of SM and the signal, and 
we observe that the $K$ factors (defined as $K=d\sigma^{NLO}/d\sigma^{LO}$) are large.
For the signal the $K$ factor is 1.98 at $Q=600~GeV$ and 1.82 at $Q=1600~GeV$. 

\begin{figure}[ht]
\centerline{\epsfig{file=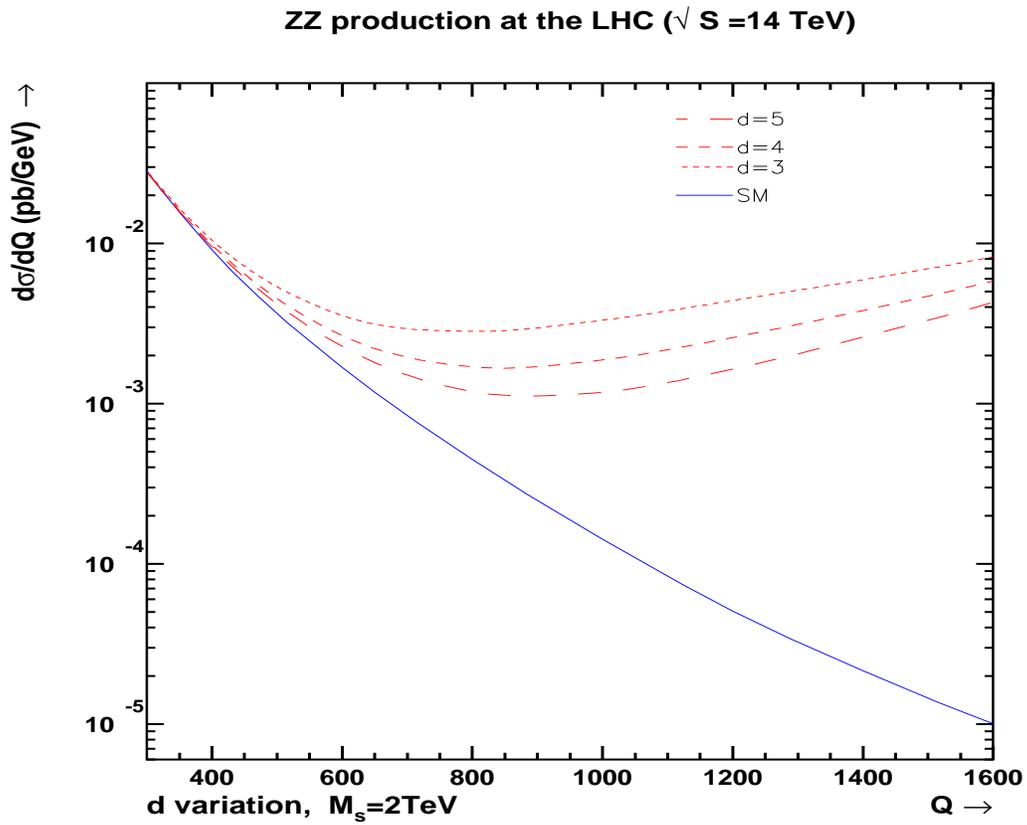,width=15cm,height=12cm,angle=0} }
\caption{
Effect of variation of number of extra dimensions in invariant mass distribution.
The fundamental scale $M_s$ has been fixed at 2 TeV.
The curves correspond to NLO results.  }
\label{invd}
\end{figure}

\begin{figure}[ht]
\centerline{\epsfig{file=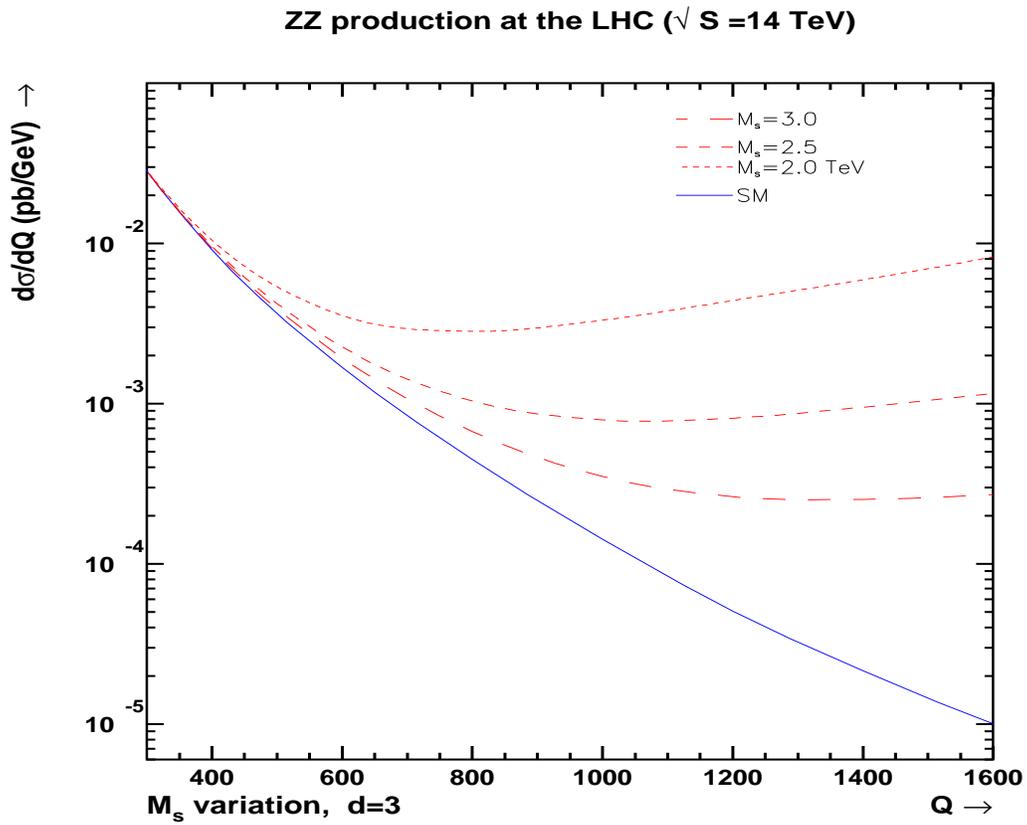,width=15cm,height=12cm,angle=0} }
\caption{Effect of variation of the fundamental scale $M_s$ in the invariant mass
distribution. The number of extra dimensions has been fixed at 3. The curves correspond
to NLO results.
}
\label{invm}
\end{figure}

To estimate the effect of the number of extra dimension on the invariant mass
distribution, we plot in Fig.~\ref{invd} the signal for three different values of $d$ (3,4,5)
with $M_s$ fixed at 2 $TeV$. We note that the lower the value of $d$ more is 
the strength of the signal. Next in Fig.~\ref{invm} we have plotted $d\sigma/dQ$ for 
three different values of $M_s$ (2.0, 2.5, 3.0) at a fixed value 3 for the  number of
extra dimensions. As expected, with increase in the fundamental scale the deviations
from SM predictions become less, and significant deviations from SM are observed 
at higher energies still. 

\begin{figure}[ht]
\centerline{\epsfig{file=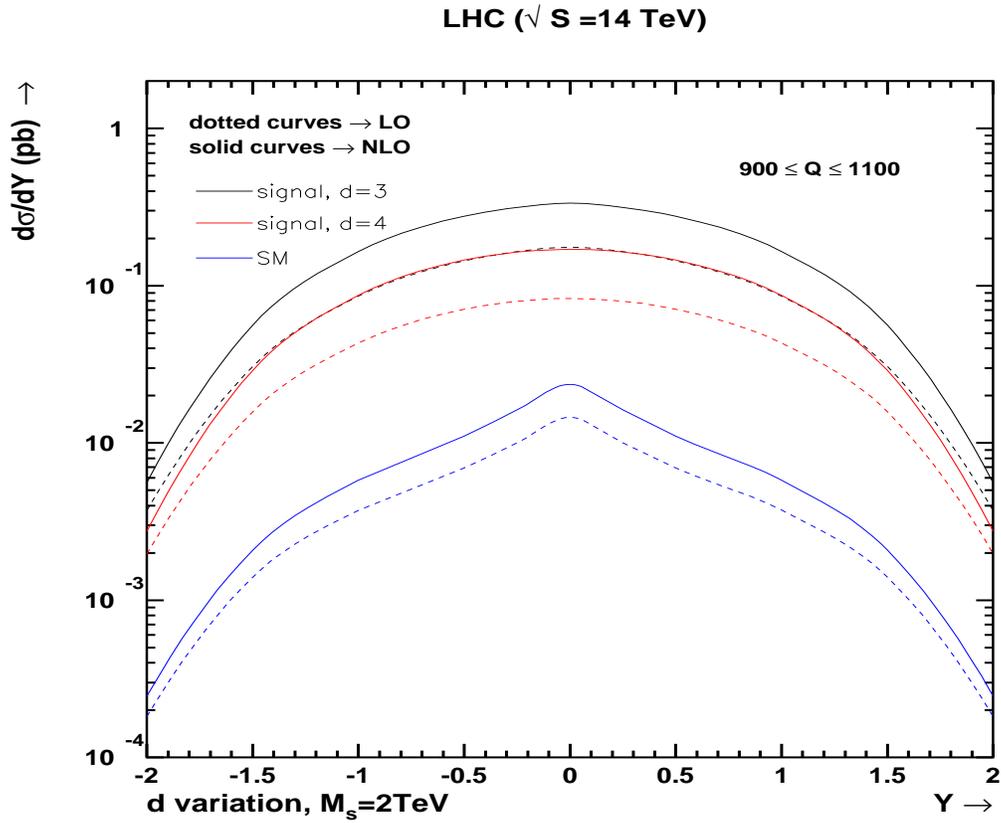,width=15cm,height=12cm,angle=0} }
\caption{
Rapidity distribution for $M_s =2TeV$ for SM and signal for $d=3$ and $d=4$. 
The dotted curves correspond to the LO and solid curves to NLO. 
We have integrated over the 
invariant mass range $900<Q<1100$ to enhance the signal.
}
\label{Y}
\end{figure}

If Fig.~\ref{Y} we have plotted the rapidity distribution $d\sigma/dY $ at LO and 
NLO both for SM and the signal for $d=3$ and $d=4$. The rapidity Y is defined as 
\be
Y =  \frac{1}{2} \ln \frac{P_1\cdot q}{P_2 \cdot q},
\ee
where $P_1$ and $P_2$ are incoming proton momenta and $q=p_3 +p_4$ ie., sum
of the $Z$ boson 4-momenta. 
We have plotted this distribution in the interval $-2.0 < Y < 2.0 $ and 
have carried out an integration over the invariant mass interval $900 < Q < 1100$
to increase the signal over the SM background.
As expected the distribution is symmetric about $Y=0$.

\begin{figure}[ht]
\centerline{\epsfig{file=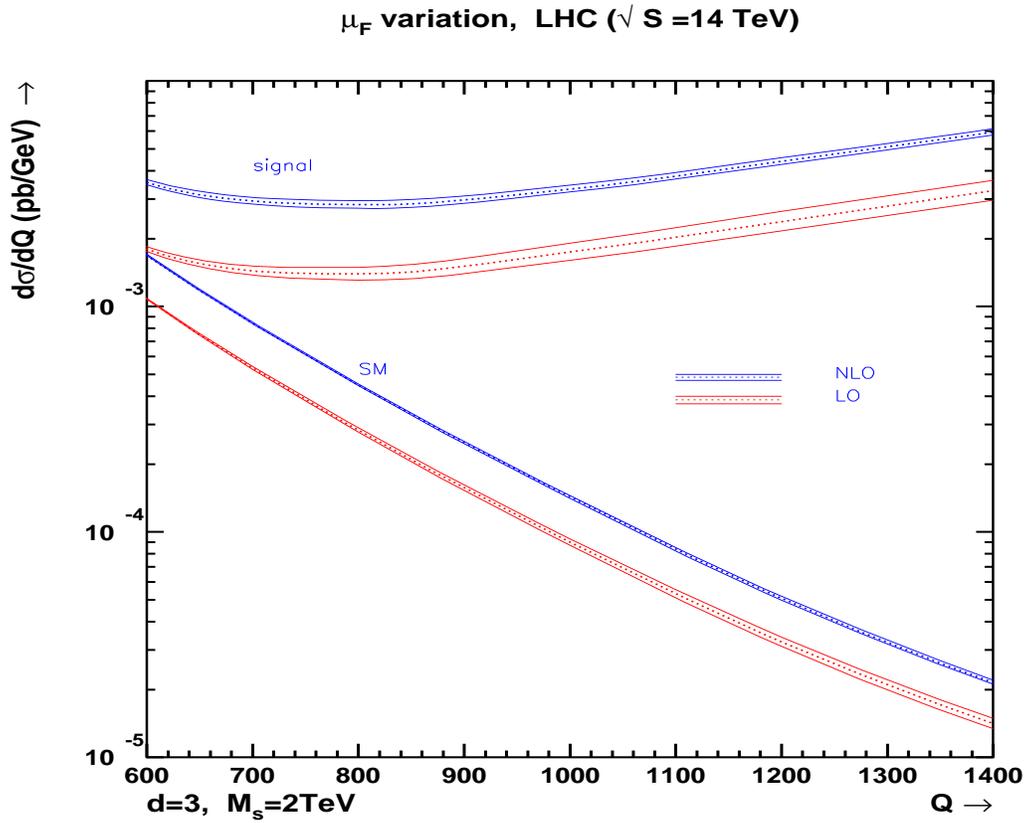,width=15cm,height=12cm,angle=0} }
\caption{Factorization scale variation in the invariant mass distribution. The number of 
extra dimensions $d=3$ and the fundamental scale $M_s=2TeV$ have been chosen. LO curves
correspond to CTEQ~6L1 density sets and NLO curves to CTEQ~6M sets.
}
\label{mufvar}
\end{figure}

We have mentioned before that the NLO QCD corrections reduce the sensitivity 
of the cross sections to the factorization scale $\mu_F$; this we now show in 
the Fig.~\ref{mufvar}. We have plotted SM and the signal both at LO and NLO,
and have varied the factorization scale $\mu_F$ in the range $Q/2 < \mu_F < 2Q$.
The central curve in a given band (shown by the dotted curves) correspond to 
$\mu_F =Q$. In all these the renormalization scale is fixed at $\mu_R =Q$.
We notice that the factorization scale uncertainties in SM are reduced compared
to the signal. 
This is because of the dominant role of the gluon gluon initiated
process in the signal. Most importantly we have been able to demonstrate that
a significant reduction in factorization scale uncertainty is achieved by
carrying out a full NLO computation.
For instance at $Q=1400~GeV$ varying $\mu_F$ between $Q/2$ to $2Q$
shows a variation of $20.4 \%$ at LO for the signal, however the NLO result at the same $Q$ value
shows a variation of $6.4\%$. 

\begin{figure}[ht]
\centerline{\epsfig{file=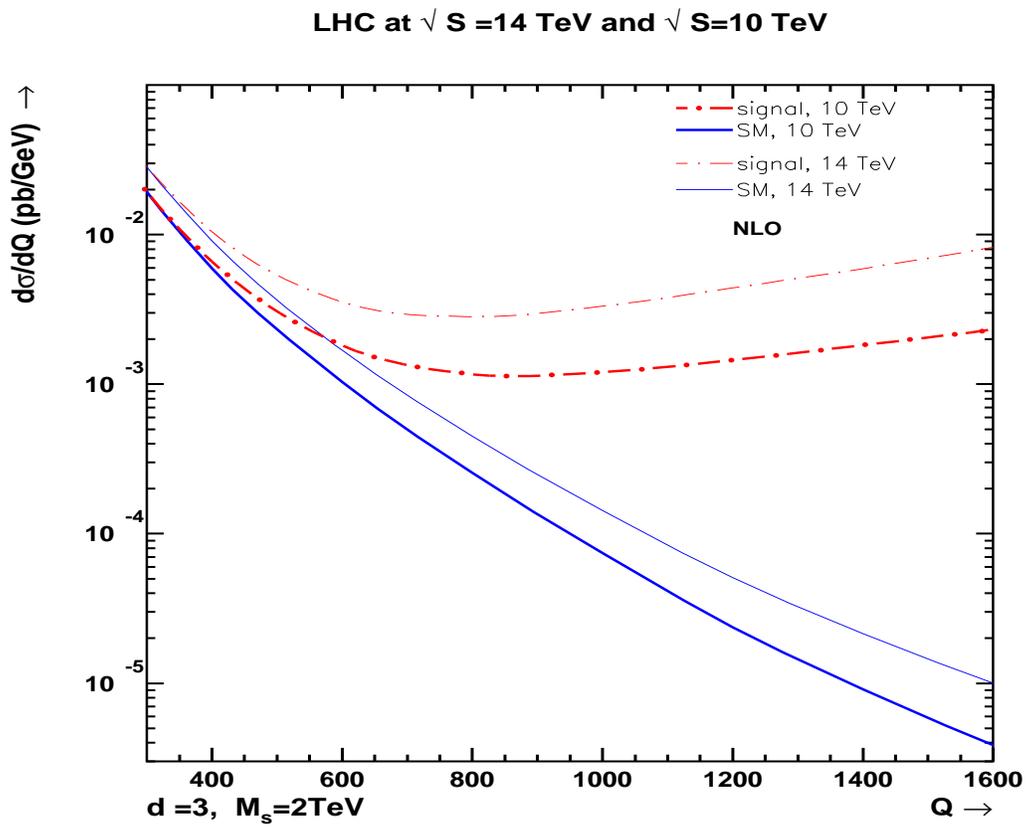,width=15cm,height=12cm,angle=0} }
\caption{
Invariant mass distribution at NLO for SM and the signal. Here the thicker
curves correspond to $\sqrt S =10 TeV$ and lighter curves to $\sqrt S=14 TeV$
at the LHC.
}
\label{ten}
\end{figure}

At the end we present in Fig.~\ref{ten}, $d\sigma/dQ$ for LHC with a centre of mass
energy of $10~TeV$ at NLO both for SM and signal. For comparison 
we have also plotted the $14~TeV$ results in the same figure.

\section{Summary and Conclusions}

In this paper we have carried out a full NLO QCD calculation for 
the production of two $Z$ bosons at the LHC at $14 TeV$ in the large
extra dimension model of ADD. Here we take all order $a_s$ contributions,
both in the SM and in the gravity mediated processes and their interferences,
into account. We have presented all the leading order and one loop virtual 
matrix element squared for the process. The method of two cutoff phase space
slicing, on which our monte carlo FORTRAN code is based, is very briefly discussed. 
After offering some checks on our monte
carlo code we 
obtained invariant mass and rapidity distributions both at LO and NLO. 
We use CTEQ 6L1 and CTEQ 6M parton density sets for LO and NLO observables, respectively.
Significant enhancements over the LO predictions are observed.
The $K$ factors are found to be large in the invariant mass distribution; for instance 
the signal has a $K$ factor of 1.98 at $Q=600~GeV$ and 1.82 at $Q=1600~GeV$. 
We have also presented the effects of variation of number of extra dimensions $d$
and the fundamental scale $M_s$ in the $Q$ distribution. We have shown that a significant
reduction in LO theoretical uncertainty, arising from the factorization scale, is achieved 
by our NLO computation. 
For instance at $Q=1400~GeV$ varying $\mu_F$ between $Q/2$ to $2Q$
shows a variation of $20.4 \%$ at LO for the signal, however the NLO result at the same $Q$ value
shows a variation of $6.4\%$. 
Thus our NLO results are more precise than the LO results
and suitable for further studies for constraining the parameters of the ADD model.
Invariant mass distribution is also presented for LHC at a center of mass energy of
$10 TeV$ at the NLO level.
\\

\noindent
{\bf Acknowledgments:}
The work of NA is supported by CSIR Senior Research Fellowship, New Delhi.
NA, AT and VR would also like to thank
the cluster computing facility at Harish-Chandra Research Institute. 
NA and VKT acknowledge the computational support of the computing facility which has been developed by
the Nuclear Particle Physics Group of the Physics Department, Allahabad University under the Center of
Advanced Study (CAS) funding of U.G.C. India.  The authors
would like to thank Prakash Mathews, Swapan Majhi and M.C. Kumar for useful discussions.




\section{Appendix}
\bea
\overline{|M^{V}|^2}^{fin}_{q \qb,sm} 
  &=& {e^4 \over N} \left( C_v^4+6~C_v^2C_a^2+C_a^4 \right) {\cal T}_W^4 
~ \left(\frac{1}{8 t^2 u^2}  \right)
\nonumber \\[2ex] 
&&\times \Bigg[
    \Bigg\{ 
           \frac{u~ F_1(t) } {(t-m^2)^2}
             \Big( 3m^8(-4t+u) + 6m^6 t (2t+u) - 2m^2t^2u (4t+u)
\nonumber \\ [2ex] 
&& 
                  +t^3u (2t+3u) + m^4t (-2t^2 +tu -3u^2) 
              \Big)      
\nonumber\\[2ex]
&&          + u~ F_2 (t) \Big( m^4(8t-2u) -4m^2t(2t+u) 
+ t(2t^2+2tu+u^2) \Big)
          + t \leftrightarrow u
     \Bigg\} 
\nonumber\\[2ex]
&&     +2\zeta(2) \Big( -20m^2tu(t+u)-4m^4(t^2-8tu+u^2) 
     +tu(5t^2+4tu+5u^2) \Big)
\nonumber \\ [2ex] 
&&
     + \frac{ F_3} {(4m^2-s)^2}
       \Big( 8m^2t^2u^2(t+u) + 12 m^8 (t^2 -8tu +u^2) 
\nonumber \\ [2ex]  
&&
       -tu(t+u)^2 (3t^2+4tu+3u^2) 
             + 4m^6(3t^3 - 5t^2u -5tu^2 +3u^3) 
\nonumber \\ [2ex] 
&&
       +m^4(3t^4 + 14t^3u +78t^2u^2 +14tu^3 +3u^4)
       \Big)
\nonumber\\[2ex]
&&     + ~\frac{t u ~F_4}{s (4m^2-s)^2}
             \Big(-32m^6tu - 64m^8(t+u)
+8m^2tu(t+u)^2
\nonumber\\[2ex]
&&            -(t+u)^3 (3t^2+4tu +3u^2)
            +m^4(22t^3 + 82t^2u +82tu^2 +22u^3) \Big)
\nonumber \\ [2ex] 
&&
     + \frac{1}{(t-m^2)(u-m^2)(4m^2-s)}
       \Big(  18m^{10} (t^2-8tu+u^2) 
\nonumber \\ [2ex]  
&&
             +m^8(-9t^3 +131 t^2u + 131 tu^2 -9u^3)
             -7t^2u^2(t^3+t^2u +tu^2 +u^3) 
\nonumber \\ [2ex]  
&&
             -4m^4 tu (4t^3 +23t^2u +23tu^2 +4u^3)
             + m^6(-9t^4 +14t^3u -66t^2u^2 
\nonumber \\ [2ex]  
&& +14tu^3 -9u^4) + m^2tu(9t^4 +32t^3u +82t^2u^2 +32tu^3 +9u^4)
       \Big)
  \Bigg]
\eea
\bea
\!\!\!\!\overline{|M^{V}|^2}^{fin}_{q \qb,int} 
  &=&  
{e^2 \kappa^2  \over 8 N} \left (C_v^2 + C_a^2 \right ) {\cal T}_W^2
\nonumber\\[2ex]
&&\times \Bigg[ \Big\{
           \frac{ G_1(t)}
                {t \left(m^2-t \right) }
           \Big( -9m^8 -4m^4t^2 + 3m^6(5t+u) -m^2tu (9t+u) 
\\ [2ex] \nonumber 
&&
+t^2u (2t+3u) \Big)
         + \frac{G_2(t)}{t} \Big( 6m^6 -2m^4(8t+u) +m^2t (10t +7u) 
\\ [2ex] \nonumber 
&&
           - t(2t^2 +2tu +u^2) \Big) + t \leftrightarrow u
    \Big\}
   -\frac{ 2 G_3}{tu} \Big( -18m^6 (t+u) 
\\ [2ex] \nonumber 
&&
-37 m^2tu (t+u) +m^4(6t^2 +80tu +6u^2) 
       +tu(7t^2 +4tu +7u^2) \Big)
\\ [2ex] \nonumber 
&&
   + \frac{G_4 }{t u \left(u-m^2 \right) \left(m^2-t \right)}
     \Big( 9m^{10} (t+u) 
+ m^4tu(t^2+10tu+u^2) 
\\ [2ex] \nonumber 
&& -6m^8\left(2t^2+5tu+2u^2\right)
          +t^2u^2(3t^2 +4tu+3u^2) 
\\ [2ex] \nonumber 
&& -m^2tu(t^3 +14t^2u +14tu^2 +u^3) 
          +m^6(3t^3+19t^2u +19tu^2 + 3u^3)
     \Big)
\\ [2ex] \nonumber 
&&   - \frac{G_5}{s \left(4m^2-s\right)^2}
     \Big( -128m^{10} +48m^8(t+u) + 48m^6(t^2+3tu +u^2) 
\\ [2ex] \nonumber 
&&
+(t+u)^3 (3t^2+4tu+3u^2) 
           -m^2(t+u)^2 (5t^2+18tu+5u^2) 
\\ [2ex] \nonumber 
&&
-2m^4(7t^3 +37t^2u +37tu^2 +7u^3) \Big)
   + \frac{G_6}{tu \left( 4m^2-s \right) }
     \Big( -114 m^8 (t+u) 
\\ [2ex] \nonumber 
&&
+m^6 (-19t^2 +334 tu -19u^2) 
-m^2tu (53t^2 +190 tu +53u^2)
\\ [2ex] \nonumber 
&&
           + 17tu ( t^3 +t^2u +tu^2 +u^3) 
+m^4 (19t^3 +61t^2u +61tu^2 +19u^3)
     \Big)
\\ [2ex] \nonumber 
&&
   + \frac{ G_7 }{ tu \left(4m^2-s \right)^2 }
     \Big(-36m^{10}(t+u) 
-8m^8(3t^2-13tu +3 u^2)
\\ [2ex] \nonumber 
&&
            +tu(t+u)^2(3t^2+4 t u +3u^2)+ m^6 (3 t^3 +61 t^2 u +61 t u^2+3 u^3)
\\ [2ex] \nonumber 
&&
            +m^4(3 t^4 -2t^3u -98t^2u^2 -2tu^3 +3u^4)
-7m^2tu(t+u)^3 
      \Big)
            \Bigg]
\eea
where
\bea
F_1  &=&\displaystyle{\ln{-\frac{t}{m^2}}}~, \quad
\quad \quad \quad
F_2    =  2 \ln \left(- \frac{t}{m^2} \right)
            \ln \left\{\frac{(t-m^2)^2}{m^2s} \right\}
            + 4 Li_2 \left( \frac{t}{m^2} \right)
            - \ln^2\left(\frac{-t}{m^2}\right)~ ,
\nonumber \\ [2ex]
F_3 &=&  \ln{\frac{s}{m^2}}~,~~
\quad \quad \quad \quad
F_4    =  \frac{1}{\beta}
            \Big[ \ln^2(\gamma) + 4 Li_2 (-\gamma) +2 \zeta(2)
              \Big] ~,~~
\nonumber \\ [2ex]
\gamma &=&  \frac{1-\beta}{1+\beta}~,
\quad \quad \quad \quad
\beta  =  \sqrt {1- 4m^2/s} .
\eea
and 
\bea
\nonumber 
G_1 &=&  \ReDs \ln \left(\frac{-t}{m^2} \right),
\\ [2ex] \nonumber 
G_2 &=&  4 \ReDs Li_2 { \left( \frac{t}{m^2} \right) }           +
         2 \ReDs \ln { \left( \frac{-t}{m^2} \right) } 
                \ln {\left( \frac{(m^2-t)^2}{m^2 s} \right) }    
\\ [2ex] \nonumber 
&&
         +2 \ImDs \pi \ln {\left( \frac{(m^2-t)^2}{m^2 s} \right)} -
           \ReDs \ln {\left( \frac{-t}{m^2} \right)^2 }- 
         2 \ImDs \pi \ln {\left( \frac{-t}{m^2} \right)},
\\ [2ex] \nonumber
G_3 &=&  \zeta(2) \ReDs,
\quad \quad \quad \quad 
G_4 =  \ImDs \pi,  
\\ [2ex] \nonumber
G_5 &=&  \frac{\ReDs}{\beta} \ln^2 {\left(\gamma\right)} +
         \frac{4}{\beta} \ReDs Li_2 {\left( -\gamma \right)} +
         \frac{2 \ReDs}{\beta} \zeta(2),
\\ [2ex] 
G_6 &=&  \ReDs ,
\quad \quad \quad \quad
G_7 =  \ReDs \ln \left(s/m^2 \right) .
\eea   

\bea
\!\!\!\! \overline{|M^{V}|^2}_{gg ,int}^{fin} 
&=& 
 \left (C_v^2 + C_a^2 \right ) {\cal T}_W^2 \frac{1}{C_A} \frac{e^2 \kappa^2}{N^2-1}
\nonumber \\ [2ex] 
&&
   \times \Bigg[
   \Big\{H_1(t) \Big(9m^4 +2t^2 +2tu +u^2 -6m^2(t+u) \Big) 
+\frac{H_2(t)}{4(t-m^2)^2} \Big(-9m^8 
\nonumber \\ [2ex] 
&&
+t^2u(2t+u) 
-2m^2tu(3t+u) 
+2m^6(5t+3u)-m^4(3t^2-2tu+u^2) \Big) 
\nonumber \\ [2ex] 
&&
+ u \leftrightarrow t \Big\}
+\frac{H_3}{4} \Big(18m^4 +3t^2 +4tu +3u^2 -12m^2(t+u) \Big)
\nonumber \\ [2ex] 
&&
+\frac{H_4}{4 (4m^2-s)^2} \Big( -32m^8 +4m^6(t+u) -(t+u)^2(t^2+4tu+u^2) 
\nonumber \\ [2ex] 
&&
+m^4(6t^2+44tu+6u^2) +2m^2(t^3-3t^2u-3tu^2+u^3) \Big)
+\frac{H_5}{s (4m^2-s)^2} \Big( 80m^{10} 
\nonumber \\ [2ex] 
&&
-32m^8(t+u) +8m^2tu(t+u)^2 -16m^6(t^2+5tu+u^2) 
\nonumber \\ [2ex] 
&&
           -(t+u)^3(3t^2+4tu+3u^2) +2m^4(5t^3 +31t^2u +31tu^2 +5u^3) \Big)
\nonumber \\ [2ex] 
&&
-\frac{H_6}{4 (t-m^2)^2 (u-m^2)^2} \Big( 18m^{12}-34m^{10}(t+u) -t^2u^2(t^2+4tu+u^2) 
\nonumber \\ [2ex] 
&&
+m^8(25t^2 +36tu+25u^2) 
            +2m^2tu(t^3+6t^2u +6tu^2+u^3) 
\nonumber \\ [2ex] 
&&
-4m^6(2t^3 +t^2u+tu^2+2u^3) +m^4(t^4-8t^3u-20t^2u^2-8tu^3+u^4) \Big)
\nonumber \\ [2ex] 
&&
+\frac{H_7}{4(t-m^2)(u-m^2)(4m^2-s)} \Big( -28m^{10} +8m^6tu +26m^8(t+u) 
\nonumber \\ [2ex] 
&&
+tu(t^3+t^2u+tu^2+u^3) -m^4(5t^3+23t^2u +23tu^2 +5u^3)
\nonumber \\ [2ex] 
&&
            + m^2(t^4 +4t^3u +10t^2u^2+4tu^3+u^4) \Big) \Bigg] 
\eea
where
\bea
\nonumber 
H_1(t) &=&  \frac{1}{8} \Bigg(  2 \ReDs \ln \left(\frac{-t}{m^2} \right) 
                               \ln \left(\frac{(m^2-t)^2}{m^2 s }\right)
\nonumber \\ [2ex] 
&&
                              + 2 \ImDs \pi \ln \left(\frac{(m^2-t)^2}{m^2 s }\right)
                              + 4 \ReDs Li_2{ \left( \frac{t}{m^2}\right)}
\nonumber \\ [2ex] 
&&
                              - \ReDs  \ln \left(\frac{-t}{m^2}\right)^2
                              -2 \ImDs \pi \ln \left(\frac{-t}{m^2}\right)
                     \Bigg),
\\ [2ex] \nonumber 
H_1(u) &=&  \frac{1}{8} \Bigg(  2 \ReDs \ln \left(\frac{-u}{m^2} \right) 
                               \ln \left(\frac{(m^2-u)^2}{m^2 s }\right)
\nonumber \\ [2ex] 
&&
                                + 2 \ImDs \pi \ln \left(\frac{(m^2-u)^2}{m^2 s }\right)
                                + 4 \ReDs Li_2{ \left( \frac{u}{m^2}\right)}
\nonumber \\ [2ex] 
&&
                                - \ReDs  \ln \left(\frac{-u}{m^2}\right)^2
                                -2 \ImDs \pi \ln \left(\frac{-u}{m^2}\right)
                     \Bigg),
\\ [2ex] \nonumber 
H_2(t) &=&   \ReDs \ln  \left( \frac{-t}{m^2} \right), 
\quad \quad \quad \quad
H_2(u) =   \ReDs \ln \left( \frac{-u}{m^2} \right), 
\\ [2ex] \nonumber 
H_3 &=&  \zeta(2) \ReDs,
\quad \quad \quad \quad 
H_4 =  \ReDs ln \left( \frac{s}{m^2} \right),
\\ [2ex] \nonumber
H_5 &=&  \frac{1}{8} \Big(  \frac{\ReDs}{\beta} \ln^2 {\left(\gamma\right)} +
                            \frac{4}{\beta} \ReDs Li_2 {\left( -\gamma \right)} +
                            \frac{2 \ReDs}{\beta} \zeta(2)
                     \Big),
\\ [2ex] \nonumber             
H_6 &=&  \ImDs \pi,
\quad \quad \quad \quad
H_7 =  \ReDs .
\eea

%
%
%
%
%
%
%
%
%
%

\vspace{1cm}
\begin{figure}[h]
\SetScale{1.5}
\noindent
\begin{center}
\hspace{2cm}
\begin{picture}(200,40)
\ArrowLine(0,40)(25,40)
\ArrowLine(25,40)(25,10)
\ArrowLine(25,10)(0,10)
\Photon(25,40)(50,40){2}{3}
\Photon(25,10)(50,10){2}{3}
\end{picture}
\caption{Leading order diagram in SM.
The diagram with the momenta of 
final state $Z$ bosons interchanged (which is not shown here) also contributes.}
\end{center}
\label{smborn}
\vspace{1.0cm}
\begin{center}
\hspace{-5cm}
\begin{picture}(200,40)
\ArrowLine(0,40)(25,40)
\ArrowLine(25,40)(25,10)
\ArrowLine(25,10)(0,10)
\Photon(25,40)(50,40){2}{3}
\Photon(25,10)(50,10){2}{3}
\GlueArc(25,25)(10,90,270){3}{3}
\ArrowLine(125,40)(150,40)
\ArrowLine(150,40)(150,10)
\ArrowLine(150,10)(125,10)
\Photon(175,10)(150,10){2}{3}
\Photon(150,40)(175,40){2}{3}
\Gluon(136,10)(136,40){2}{3}
\Text(35,1)[]{(a)}
\Text(235,1)[]{(b)}
\end{picture}
\end{center}
\vspace{1.0cm}
\begin{center}
\hspace{-5cm}
\begin{picture}(200,40)
\ArrowLine(0,40)(25,40)
\ArrowLine(25,40)(25,10)
\ArrowLine(25,10)(0,10)
\Photon(25,40)(50,40){2}{3}
\Photon(25,10)(50,10){2}{3}
\GlueArc(25,10)(15,90,180){2}{3}
\ArrowLine(125,40)(150,40)
\ArrowLine(150,40)(150,10)
\ArrowLine(150,10)(125,10)
\Photon(175,10)(150,10){2}{3}
\Photon(150,40)(175,40){2}{3}
\GlueArc(150,40)(15,180,270){2}{3}
\Text(35,1)[]{(c)}
\Text(235,1)[]{(d)}
\end{picture}
\end{center}
\vspace{1.0cm}
\begin{center}
\hspace{-5cm}
\begin{picture}(200,40)
\ArrowLine(15,40)(40,40)
\ArrowLine(40,40)(40,10)
\ArrowLine(40,10)(15,10)
\ArrowLine(15,10)(15,40)
\Photon(40,40)(55,40){2}{3}
\Photon(40,10)(55,10){2}{3}
\Gluon(0,40)(15,40){2}{3}
\Gluon(0,10)(15,10){2}{3}
\Text(35,1)[]{(e)}
\ArrowLine(140,10)(140,40)
\ArrowLine(140,40)(165,10)
\ArrowLine(165,10)(165,40)
\Line(165,40)(140,10)
\Photon(165,10)(180,10){2}{3}
\Photon(180,40)(165,40){2}{3}
\Gluon(125,40)(140,40){2}{3}
\Gluon(125,10)(140,10){2}{3}
\Text(235,1)[]{(f)}
\end{picture}
\end{center}
\vspace{1.0cm}
\begin{center}
\hspace{-5cm}
\begin{picture}(200,40)
\ArrowLine(15,40)(40,40)
\ArrowLine(40,40)(15,10)
\ArrowLine(15,10)(40,10)
\Line(40,10)(15,40)
\Photon(40,40)(55,40){2}{3}
\Photon(40,10)(55,10){2}{3}
\Gluon(0,40)(15,40){2}{3}
\Gluon(0,10)(15,10){2}{3}
\Text(35,1)[]{(g)}
\end{picture}
\end{center}
\caption{Order $a_s$ virtual diagrams in SM.
The diagrams with the momenta of 
final state $Z$ bosons interchanged (which are not shown here) also contribute.}
\label{smvrt}
\end{figure}

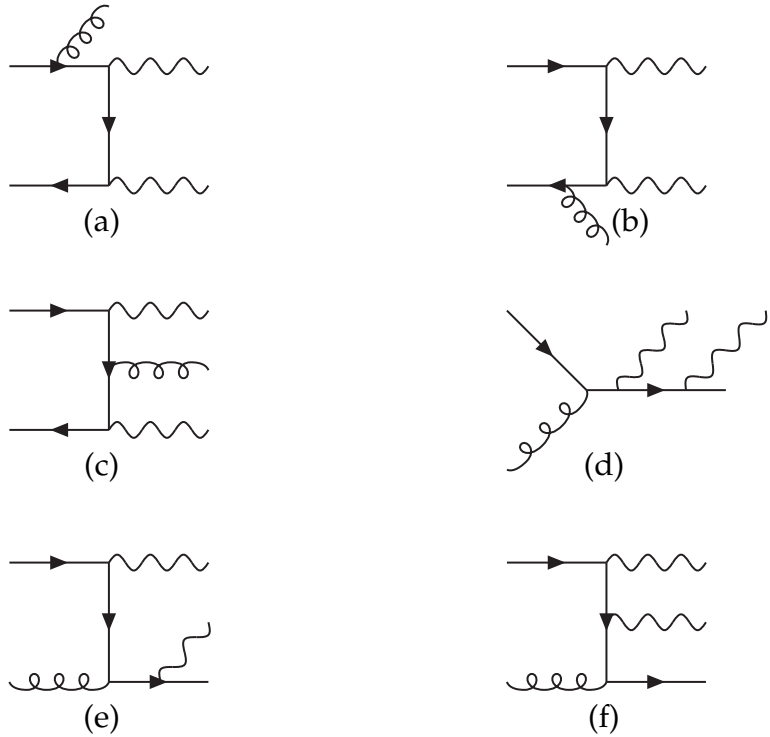
\begin{figure}
\SetScale{1.5}
\noindent
\begin{center}
\hspace{-5cm}
\begin{picture}(200,40)
\ArrowLine(0,40)(25,40)
\ArrowLine(25,40)(25,10)
\ArrowLine(25,10)(0,10)
\Photon(25,40)(50,40){2}{3}
\Photon(25,10)(50,10){2}{3}
\Gluon(12.5,40)(25,55){2}{3}
\Text(35,1)[]{(a)}
\ArrowLine(125,40)(150,40)
\ArrowLine(150,40)(150,10)
\ArrowLine(150,10)(125,10)
\Photon(175,10)(150,10){2}{3}
\Photon(150,40)(175,40){2}{3}
\Gluon(137.5,10)(150,-5){2}{3}
\Text(235,1)[]{(b)}
\end{picture}
\end{center}
\vspace{1.0cm}
\begin{center}
\hspace{-5cm}
\begin{picture}(200,40)
\ArrowLine(0,40)(25,40)
\ArrowLine(25,40)(25,10)
\ArrowLine(25,10)(0,10)
\Photon(25,40)(50,40){2}{3}
\Photon(25,10)(50,10){2}{3}
\Gluon(25,25)(50,25){2}{3}
\Text(35,1)[]{(c)}
\ArrowLine(125,40)(145,20)
\Gluon(145,20)(125,0){2}{3}
\ArrowLine(145,20)(180,20)
\Photon(153,20)(170,40){2}{3}
\Photon(170,20)(190,40){2}{3}
\Text(225,1)[]{(d)}
\end{picture}
\end{center}
\vspace{1.0cm}
\begin{center}
\hspace{-5cm}
\begin{picture}(200,40)
\ArrowLine(0,40)(25,40)
\ArrowLine(25,40)(25,10)
\Gluon(25,10)(0,10){2}{3}
\ArrowLine(25,10)(50,10)
\Photon(25,40)(50,40){2}{3}
\Photon(38,10)(50,25){2}{2}
\Text(35,1)[]{(e)}
\ArrowLine(125,40)(150,40)
\ArrowLine(150,40)(150,10)
\Gluon(150,10)(125,10){2}{3}
\ArrowLine(150,10)(175,10)
\Photon(175,25)(150,25){2}{3}
\Photon(150,40)(175,40){2}{3}
\Text(225,1)[]{(f)}
\vspace{1.0cm}
\end{picture} 
\end{center}
\vspace{1.0cm}
\vspace{1.0cm}
\caption{Order $a_s$ real emission Feynman diagrams in SM.
The diagrams with the momenta of 
final state $Z$ bosons interchanged (which are not shown here) also contribute.}
\label{smreal}
\end{figure}

\begin{figure}[h]
\SetScale{1.5}
\noindent
\begin{center}
\hspace{-5cm}
\begin{picture}(200,40)
\ArrowLine(0,40)(25,20)
\ArrowLine(25,20)(0,0) 
\DashLine(25,20)(50,20){4}
\Photon(50,20)(75,40){2}{3}
\Photon(50,20)(75,0){2}{3}
\Text(65,1)[]{(a)}
\Gluon(125,40)(150,20){2}{3}
\Gluon(150,20)(125,0){2}{3}
\DashLine(150,20)(175,20){4}
\Photon(175,20)(200,40){2}{3}
\Photon(175,20)(200,0){2}{3}
\Text(245,1)[]{(b)}
\end{picture}
\end{center}
\caption{LO gravity mediated diagrams.} 
\label{bsmborn}
\vspace{2cm}
\SetScale{1.5}
\noindent
\begin{center}
\hspace{-5cm}
\begin{picture}(200,40)
\ArrowLine(0,40)(20,40)
\ArrowLine(20,40)(20,0)
\ArrowLine(20,0)(0,0)
\Gluon(20,40)(38,20){2}{3}
\Gluon(38,20)(20,0){2}{3}
\DashLine(38,20)(55,20){4}
\Photon(55,20)(75,40){2}{3}
\Photon(55,20)(75,0){2}{3}
\Text(65,1)[]{(a)}
\ArrowLine(125,40)(150,20)
\ArrowLine(150,20)(125,0)
\Gluon(133,8)(133,33){2}{3}
\DashLine(150,20)(175,20){4}
\Photon(175,20)(200,40){2}{3}
\Photon(175,20)(200,0){2}{3}
\Text(245,1)[]{(b)}
\end{picture}
\end{center}
%
\vspace{1.0cm}
\begin{center}
\hspace{-5cm}
\begin{picture}(200,40)
\Gluon(0,40)(20,40){2}{3}
\ArrowLine(20,0)(20,40)
\Gluon(20,0)(0,0){2}{3}
\ArrowLine(20,40)(38,20)
\ArrowLine(38,20)(20,0)
\DashLine(38,20)(55,20){4}
\Photon(55,20)(75,40){2}{3}
\Photon(55,20)(75,0){2}{3}
\Text(65,1)[]{(c)}
\Gluon(125,40)(145,40){2}{3}
\ArrowLine(145,40)(145,0)
\Gluon(145,0)(125,0){2}{3}
\ArrowLine(163,20)(145,40)
\ArrowLine(145,0)(163,20)
\DashLine(163,20)(180,20){4}
\Photon(180,20)(200,40){2}{3}
\Photon(180,20)(200,0){2}{3}
\Text(245,1)[]{(d)}
\end{picture}
\end{center}
\vspace{1.0cm}
\begin{center}
\hspace{-5cm}
\begin{picture}(200,40)
\Gluon(0,40)(20,40){2}{3}
\DashArrowLine(20,0)(20,40){1}
\Gluon(20,0)(0,0){2}{3}
\DashArrowLine(20,40)(38,20){1}
\DashArrowLine(38,20)(20,0){1}
\DashLine(38,20)(55,20){4}
\Photon(55,20)(75,40){2}{3}
\Photon(55,20)(75,0){2}{3}
\Text(65,1)[]{(e)}
\Gluon(125,40)(145,40){2}{3}
\DashArrowLine(145,40)(145,0){1}
\Gluon(145,0)(125,0){2}{3}
\DashArrowLine(163,20)(145,40){1}
\DashArrowLine(145,0)(163,20){1}
\DashLine(163,20)(180,20){4}
\Photon(180,20)(200,40){2}{3}
\Photon(180,20)(200,0){2}{3}
\Text(245,1)[]{(f)}
\end{picture}
\end{center}
\vspace{1.0cm}
\begin{center}
\hspace{-5cm}
\begin{picture}(200,40)
\Gluon(0,40)(20,20){2}{3}
\Gluon(20,20)(0,0){2}{3}
\GlueArc(29,20)(9,0,360){2}{8}
\DashLine(38,20)(55,20){4}
\Photon(55,20)(75,40){2}{3}
\Photon(55,20)(75,0){2}{3}
\Text(65,1)[]{(g)}
\Gluon(125,40)(145,40){2}{3}
\Gluon(145,0)(145,40){2}{4}
\Gluon(145,0)(125,0){2}{3}
\Gluon(145,40)(163,20){2}{3}
\Gluon(163,20)(145,0){2}{3}
\DashLine(163,20)(180,20){4}
\Photon(200,40)(180,20){2}{3}
\Photon(180,20)(200,0){2}{3}
\Text(245,1)[]{(h)}
\end{picture}
\end{center}
\caption{Order $a_s$ gravity mediated virtual correction
Feynman diagrams.}
\label{bsmvrt}
\end{figure}

\begin{figure}
\SetScale{1.5}
\noindent
\vspace{2.0cm}
\begin{center}
\hspace{-5cm}
\begin{picture}(200,40)
\ArrowLine(0,40)(25,40)
\ArrowLine(25,40)(25,10)
\ArrowLine(25,10)(0,10)
\Gluon(25,40)(50,40){2}{3}
\DashLine(25,10)(38,10){4}
\Photon(38,10)(50,20){2}{2}
\Photon(38,10)(50,0){2}{2}
\Text(45,1)[]{(a)}
\ArrowLine(125,40)(150,40)
\ArrowLine(150,40)(150,10)
\ArrowLine(150,10)(125,10)
\Gluon(175,10)(150,10){2}{3}
\DashLine(150,40)(163,40){4}
\Photon(163,40)(175,50){2}{2}
\Photon(163,40)(175,30){2}{2}
\Text(245,1)[]{(b)}

\end{picture}
\end{center}
\vspace{1.0cm}
\begin{center}
\hspace{-5cm}
\begin{picture}(200,40)
\ArrowLine(0,40)(25,20)
\ArrowLine(25,20)(0,0)
\Gluon(25,20)(50,40){2}{3}
\DashLine(25,20)(50,20){4}
\Photon(50,20)(75,40){2}{3}
\Photon(50,20)(75,0){2}{3}
\Text(45,1)[]{(c)}
\ArrowLine(125,40)(145,20)
\ArrowLine(145,20)(125,0)
\Gluon(145,20)(163,20){2}{4}
\Gluon(163,20)(180,40){2}{4}
\DashLine(163,20)(180,20){4}
\Photon(180,20)(200,40){2}{3}
\Photon(180,20)(200,0){2}{3}
\Text(245,1)[]{(d)}
\end{picture}
\end{center}
\vspace{1.0cm}
\begin{center}
\hspace{-5cm}
\begin{picture}(200,40)
\ArrowLine(0,40)(25,40)
\ArrowLine(25,40)(25,10)
\Gluon(25,10)(0,10){2}{3}
\ArrowLine(50,10)(25,10)
\DashLine(25,40)(38,40){4}
\Photon(38,40)(50,50){2}{2}
\Photon(38,40)(50,30){2}{2}
\Text(45,1)[]{(e)}
\ArrowLine(125,40)(145,20)
\Gluon(145,20)(125,0){2}{3}
\ArrowLine(145,20)(163,20)
\ArrowLine(163,20)(180,40)
\DashLine(163,20)(180,20){4}
\Photon(180,20)(200,40){2}{3}
\Photon(180,20)(200,0){2}{3}
\Text(245,1)[]{(f)}
\end{picture}
\end{center}
\vspace{1.0cm}
\begin{center}
\hspace{-5cm}
\begin{picture}(200,40)
\ArrowLine(0,40)(25,40)
\Gluon(25,40)(25,10){2}{3}
\Gluon(25,10)(0,10){2}{3}
\ArrowLine(25,40)(50,40)
\DashLine(25,10)(38,10){4}
\Photon(38,10)(50,20){2}{2}
\Photon(38,10)(50,0){2}{2}
\Text(45,1)[]{(g)}
\ArrowLine(125,40)(150,20)
\Gluon(150,20)(125,0){2}{3}
\ArrowLine(150,20)(175,40)
\DashLine(150,20)(175,20){4}
\Photon(175,20)(200,40){2}{3}
\Photon(175,20)(200,0){2}{3}
\Text(245,1)[]{(h)}
\end{picture}
\end{center}
\vspace{1.0cm}
\begin{center}
\hspace{-5cm}
\begin{picture}(200,40)
\Gluon(0,40)(25,40){2}{3}
\Gluon(25,40)(25,10){2}{3}
\Gluon(25,10)(0,10){2}{3}
\Gluon(25,40)(50,40){2}{3}
\DashLine(25,10)(38,10){4}
\Photon(38,10)(50,20){2}{2}
\Photon(38,10)(50,0){2}{2}
\Text(45,1)[]{(i)}
\Gluon(125,40)(150,40){2}{3}
\Gluon(150,40)(150,10){2}{3}
\Gluon(150,10)(125,10){2}{3}
\Gluon(175,10)(150,10){2}{3}
\DashLine(150,40)(163,40){4}
\Photon(163,40)(175,50){2}{2}
\Photon(163,40)(175,30){2}{2}
\Text(245,1)[]{(j)}
\end{picture}
\end{center}
\vspace{1.0cm}
\begin{center}
\hspace{-5cm}
\begin{picture}(200,40)
\Gluon(0,40)(25,20){2}{3}
\Gluon(25,20)(0,0){2}{3}
\Gluon(25,20)(50,40){2}{3}
\DashLine(25,20)(50,20){4}
\Photon(50,20)(75,40){2}{3}
\Photon(50,20)(75,0){2}{3}
\Text(45,1)[]{(k)}
\Gluon(125,40)(145,20){2}{3}
\Gluon(145,20)(125,0){2}{3}
\Gluon(145,20)(163,20){2}{4}
\Gluon(163,20)(180,40){2}{4}
\DashLine(163,20)(180,20){4}
\Photon(180,20)(200,40){2}{3}
\Photon(180,20)(200,0){2}{3}
\Text(245,1)[]{(l)}
\end{picture}
\end{center}
\caption{Gravity mediated real emission diagrams which
contribute at NLO.}
\label{bsmreal}
\end{figure}
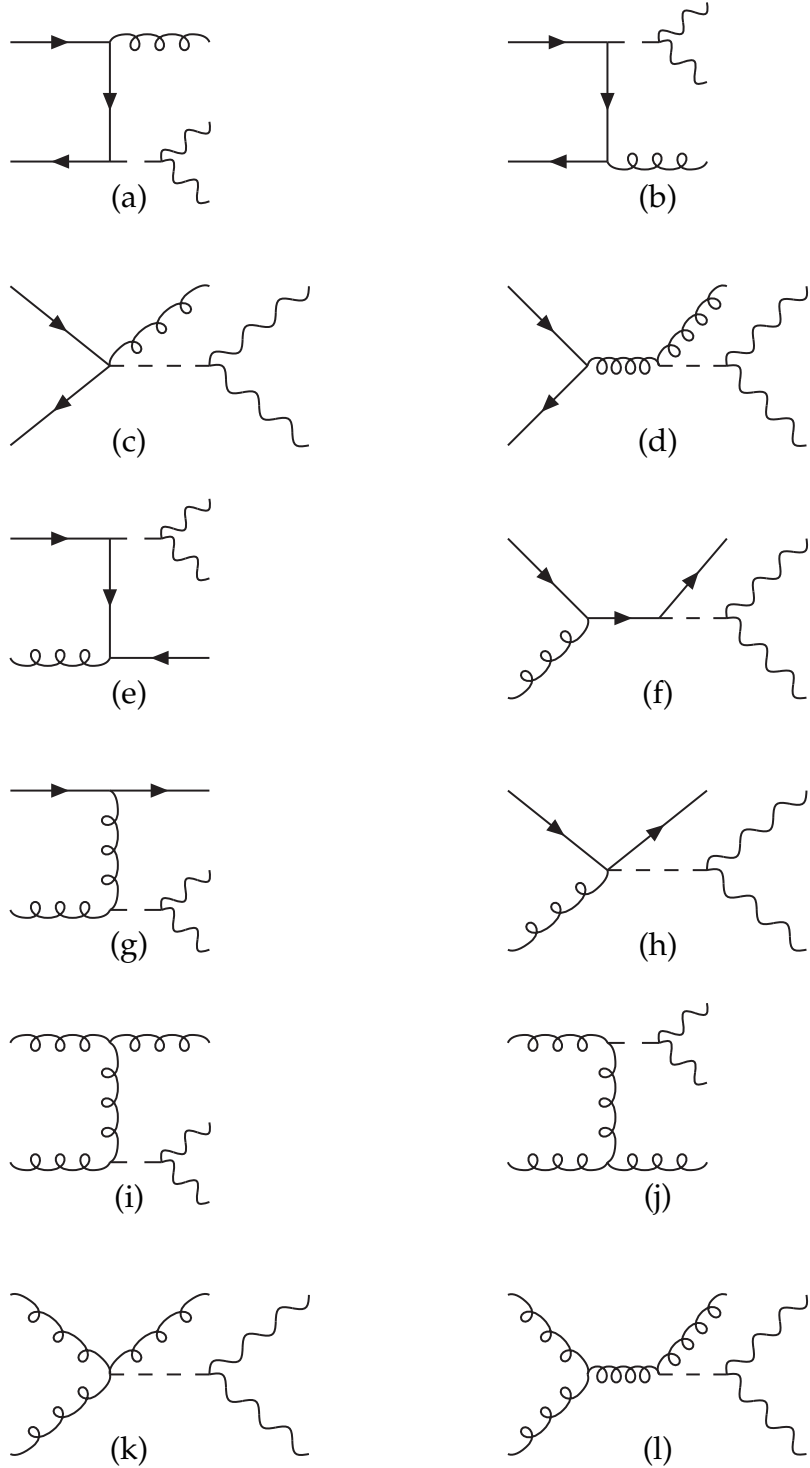

\end{document}